\documentclass[preprint]{aastex631}

\usepackage{multirow}
\usepackage{booktabs}
\usepackage{mathrsfs}
\usepackage{amsmath}
\usepackage{graphicx}
\usepackage{color}
\usepackage{dcolumn}
\newcolumntype{d}[1]{D{.}{.}{#1}}

\graphicspath{{./}{figures/}}

\begin{document}

\title{Effects of $\phi$-meson on properties of hyperon stars in density dependent relativistic mean field model}

\author[0000-0001-6836-9339]{Zhong-Hao Tu}
\affiliation{CAS Key Laboratory of Theoretical Physics,~Institute of Theoretical Physics,~Chinese Academy of Sciences,~Beijing,~100190,~China}
\affiliation{School of Physical Sciences,~University of Chinese Academy of Sciences,~Beijing,~100049,~China}

\author[0000-0003-4753-3325]{Shan-Gui Zhou}
\affiliation{CAS Key Laboratory of Theoretical Physics,~Institute of Theoretical Physics,~Chinese Academy of Sciences,~Beijing,~100190,~China}
\affiliation{School of Physical Sciences,~University of Chinese Academy of Sciences,~Beijing,~100049,~China}
\affiliation{Center of Theoretical Nuclear Physics,~National Laboratory of Heavy Ion Accelerator,~Lanzhou,~730000,~China}
\affiliation{Synergetic Innovation Center for Quantum Effects and Application,~Hunan Normal University,~Changsha,~410081,~China}

\correspondingauthor{Shan-Gui Zhou}
\email{sgzhou@itp.ac.cn}



\begin{abstract}

\noindent The effects of~$\phi$-meson on properties of hyperon stars are studied systematically in the framework of the density dependent relativistic mean field~(DDRMF)~model. The~$\phi$-meson shifts hyperon threshold to a higher density and reduces the hyperon fractions in neutron star cores. It also strongly stiffens the equation of state (EoS) calculated with various DDRMF effective interactions and increases the maximum mass of hyperon stars, but only a few effective interactions survive under the constraints from recent astrophysical observations. In the DDRMF model, the conformal limit of sound velocity is still in a strong tension with the fact that the maximum mass of neutron stars obtained in theoretical calculations reaches about two solar masses. Based on different interior composition assumptions, we discuss the possibility of the secondary object of GW190814~as a neutron star. When~$\phi$-meson is considered, DD-ME2 and DD-MEX support that the secondary object of GW190814~is a hyperon star rapidly rotating with Kepler frequency.

\end{abstract}



\section{Introduction} \label{sec:intro}
Neutron stars provide an ideal laboratory to probe the physical mechanism of dense matter at baryon number density~$\rho_{B}$~above two times the nuclear saturation density~$\rho_{0}$~and isospin asymmetry close to pure neutron matter. The interior composition and equation of state (EoS) of neutron star cores ($\rho_{B}\gtrsim 2\rho_{0}$) are still little known. Based on different theoretical assumptions, a number of possibilities have been proposed for the composition of the inner core, such as nucleons~\citep{Zhu2019_PRC99-025804,Zhu2018_PRC97-035805,Zhu2018_ApJ862-98,Li2016_ApJSupp223-16}, nucleons mixed with excited nucleons~($\Delta$)~\citep{Xiang2003_PRC67-038801,Drago2014_PRC90-065809,Zhu2016_PRC94-045803,Li2018_PLB783-234,Sun2019_PRD99-023004,Thapa2021_MNRAS507-2991,Thapa2021_PRD103-063004}, strange meson condensation~\citep{Schaffner1996_PRC53-1416,Li2006_PRC74-055801,Li2007_CP16-1934,Li2010_PRC81-025806,Lim2014_PRC89-055804,Thapa2020_PRD102-123007,Thapa2021_PRD103-063004}, deconfined quarks (hybrid stars and quark stars)~\citep{Shao2013_PRD87-096012,Alford2013_PRD88-083013,Wei2017_PRD96-043008,Maslov2019_PRC100-025802,Xia2016_PRD93-085025,Xia2016_SB61-172}, and hyperons (hyperon stars)~\citep{Glendenning1985_ApJ293-470,Schaffner1996_PRC53-1416,Li2007_CP16-1934,Weissenborn2012_PRC85-065802,Oertel2015_JPG42-075202,Katayama2015_PLB747-43,Providencia2019_FASS6-13,Hong2019_CTP71-819,Thapa2021_PRD103-063004,Rather2021_ApJ917-46}. The different composition assumptions of neutron star cores lead to large uncertainties in the EoS of neutron stars. Exploring the interior composition and constraining the EoS require combining astrophysics and nuclear physics.

Several neutron stars with about two solar masses, including PSR J1614-2230 with 1.908$\pm$0.016 $M_{\odot}$ \citep{Demorest2010_Nature467-1081}, PSR J0348+0432 with 2.01$\pm$0.04 $M_{\odot}$ \citep{Antoniadis2013_Science340-1233232}, and MSP J0740+6620 with 2.08$_{-0.07}^{+0.07} M_{\odot}$ \citep{Fonseca2021_arXiv2104.00880}, put a strong constraint on the maximum mass of neutron stars calculated with theoretical approaches. The first gravitational wave (GW) signal from GW170817 of binary neutron star (BNS) merger was observed by the LIGO Scientific Collaboration and Virgo Collaboration~\citep{Abbott2017_PRL119-161101}, and the constraints from the GW signal on the EoS and neutron star radius were performed in~\citet{Abbott2018_PRL121-161101}. The observation of GW170817~opened up a new era of multimessenger astronomy. The Neutron star Interior Composition Explorer (NICER) has been devoted to the study of the internal structure of neutron stars by soft X-ray timing~\citep{Gendreau2017_NatAstron1-895}, and a large number of studies have been performed to constrain the EoS and mass--radius relation of neutron stars~\citep{Miller2019_ApJL887-L24,Bogdanov2019_ApJL887-L25,Bogdanov2019_ApJL887-L26,Raaijmakers2020_ApJL893-L21,Jiang2020_ApJ892-55}. The advanced astronomical observation techniques will provide more information about neutron stars and help us to investigate the interior compositions of neutron stars and the behaviors of EoS of dense matter at extreme densities.

Various theoretical approaches have been used to study the properties of homogeneous nuclear matter, including \emph{ab initio}~\citep{Zuo2002_EPJA14-469,Zuo2002_NPA706-418,Baldo2004_PRC69-014301,Dickhoff2004_PPNP52-377,Bombaci2005_PLB609-232,Lee2009_PPNP63-117,Baldo2012_PRC86-064001,Carlson2015_RMP87-1067} and phenomenological~\citep{Glendenning1985_ApJ293-470,RikovskaStone2003_PRC68-034324,Stone2007_PPNP58-587,Dutra2012_PRC85-035201,Sellahewa2014_PRC90-054327,Dutra2014_PRC90-055203,Whittenbury2014_PRC89-065801}~approaches, and these studies cover large ranges of baryon number density, temperature, and isospin asymmetry. The relativistic mean field (RMF) model, one of the phenomenological models, provides an excellent tool to study the properties of infinite nuclear matter ~\citep{Chin1974_PLB52-24,Walecka1974_APNY83-491,Glendenning1996_CompactStars,Shen1999_CTP31-153,Shen2002_PRC65-035802,Shen2011_ApJSupp197-20,Zhao2012_PRC85-065806,Zhao2014_EPJA50-80,Dutra2014_PRC90-055203,Meng2016_RDFNS-21,Mu2017_ApJ846-140,Bhuyan2017_IJMPE26-1750052,Biswal2019_ApJ885-25,Tong2020_PRC101-035802}. In the framework of the RMF model, the baryons are treated as point particles and interact with each other through the exchange of scalar and vector mesons~\citep{Walecka1974_APNY83-491,Glendenning1985_ApJ293-470}. The coupling constants between nucleons and mesons are determined by fitting nuclear matter properties and/or the properties of selected finite nuclei~\citep{Schaffner1996_PRC53-1416}. Using different parameterization strategies, like non-linear (NL) and density dependent (DD), EoSs of nuclear matter with different stiffness are obtained at high density even though they are constrained well at saturation density. A RMF effective interaction may be excluded if its predictions are incompatible with astrophysical observations.

As we mentioned above, various hyperons may be populated in the inner core of neutron stars at density of about~2--3$\rho_{0}$~once the nucleon Fermi energy reaches the (in-medium) rest masses of hyperons~\citep{Glendenning1985_ApJ293-470}. However, the appearance of hyperons results in the \emph{hyperon puzzle}: Hyperons strongly soften the EoS so that the maximum mass is not compatible with observations~\citep{Schulze2006_PRC73-058801, Vidana2013_NPA914-367}. The solution of the \emph{hyperon puzzle} requires additional repulsive interaction between baryons~\citep{Vidana2013_NPA914-367}~and mechanisms that provide such repulsion include: (a)~Addition of the repulsive hyperonic three-body force~\citep{Lonardoni2015_PRL114-092301,Wirth2016_PRL117-182501}; (b)~a deconfinement phase transition to quark matter below the hyperon thresholds \citep{Weissenborn2011_ApJL740-L14,Bonanno2012_AA539-A16,Klaehn2013_PRD88-085001}; (c)~addition of the repulsive hyperonic interaction by exchanging vector meson~\citep{Weissenborn2012_PRC85-065802,Maslov2015_PLB748-369,Bhuyan2017_IJMPE26-1750052}. In the framework of the RMF model, the additional repulsive interaction between hyperons can be achieved by including the strange meson $\phi$~\citep{Schaffner1996_PRC53-1416,Weissenborn2013_NPA914-421,Zhao2015_PRC92-055802,Fortin2017_PRC95-065803,Tolos2017_PASA34-e065,Lopes2021_NPA1009-122171}.

Effects of~$\phi$-meson on properties of neutron stars have been studied in the NLRMF model~\citep{Weissenborn2012_NPA881-62,Banik2014_ApJSupp214-22,Biswal2019_ApJ885-25,Lopes2020_EPJA56-122}. But in the DDRMF model, the effects of~$\phi$-meson on the interior composition, EoS, sound velocity, and mass--radius relation of neutron stars are still missing. In the present work, combining with the widely used effective interactions DD-ME2 \citep{Lalazissis2005_PRC71-024312}, DD-ME$\delta$ \citep{Roca-Maza2011_PRC84-054309}, PKDD \citep{Long2004_PRC69-034319}, and TW99 \citep{Typel1999_NPA656-331} and the latest proposed effective interactions DD-MEX \citep{Taninah2020_PLB800-135065}, DDV, DDVT, and DDVTD \citep{Typel2020_EPJA56-160}, we systematically study the effects of~$\phi$-meson on properties of hyperon stars in the framework of the DDRMF model. The recent astrophysical observations, e.g., MSP J0740+6620 and GW190814 event, are used to discuss the significance of~$\phi$-meson on hyperon stars. For studying the effects of $\phi$-meson solely, the other strange meson $\sigma^{*}$, whose properties are little known, is neglected in this work.

This paper is organized as follows. In Sec.~\ref{sec:theor}, the theoretical framework of the DDRMF model and the methodology for calculating mass--radius relation of neutron stars are given. In Sec.~\ref{sec:resul}, we present the results and discussions on the interior compositions, EoS, sound velocity, and mass--radius relation of neutron stars. Finally, a brief summary is given in Sec.~\ref{sec:summar}.
\section{Theoretical Framework} \label{sec:theor}
Since the Walecka model was proposed \citep{Walecka1974_APNY83-491}, the RMF approach has been extended to the NLRMF and DDRMF models in order to simulate the medium-dependent effective interaction~\citep{Boguta1977_NPA292-413,Reinhard1989_RPP52-439,Fuchs1995_PRC52-3043}, and both have been successfully applied to the study of nuclear matter and finite nuclei~\citep{Reinhard1989_RPP52-439,Ring1996_PPNP37-193,Bender2003_RMP75-121,Niksic2011_PPNP66-519,Meng2015_JPG42-093101,Oertel2017_RMP89-015007,Meng2006_PPNP57-470}. Unlike the NLRMF model, the coupling constants between baryons and mesons in the DDRMF model are density dependent. In the present work, we focus on the DDRMF model and consider all octet baryons ($n$,~$p$,~$\Lambda$,~$\Sigma^{+}$,~$\Sigma^{0}$,~$\Sigma^{-}$,~$\Xi^{0}$, and~$\Xi^{-}$) and isoscalar mesons~($\sigma$~and~$\omega$), isovector mesons~($\rho$~and~$\delta$), and the hidden-strangeness vector meson~$\phi$. For convenience, we use the $npe\mu$ matter to represent the nuclear matter that consists of nucleons, leptons, and non-strangeness mesons. If hyperons are included as well, we call it the~$npe\mu Y$~matter. In the~$npe\mu Y\phi$~matter the~$\phi$-meson effects are taken into account. The corresponding neutron stars are labeled as~$npe\mu$, $npe\mu Y$, and~$npe\mu Y\phi$ neutron stars, respectively. The general Lagrangian of the DDRMF model that describes infinite nuclear matter of neutron star cores can be written as
\begin{equation}\label{equ:DDRMF_Lagrangian}
\begin{aligned}
    \mathcal{L} = & \sum_{B}\bar{\psi}_{B}\left\{ \gamma^{\mu}\left[ i\partial_{\mu}-\Gamma_{\omega B}(\rho_{B})\omega_{\mu}
                 -\Gamma_{\rho B}(\rho_{B})\boldsymbol{\rho}_{\mu}\boldsymbol{\tau}_{B}-\Gamma_{\phi B}(\rho_{B})\phi_{\mu} \right]
                 -\left[ M_{B}+\Gamma_{\sigma B}(\rho_{B})\sigma+\Gamma_{\delta B}(\rho_{B})\boldsymbol{\delta}\boldsymbol{\tau}_{B} \right] \right\}\psi_{B} \\
                &+\frac{1}{2}(\partial^{\mu}\sigma\partial_{\mu}\sigma-m_{\sigma}^{2}\sigma^{2})
                 +\frac{1}{2}(\partial^{\mu}\boldsymbol{\delta}\partial_{\mu}\boldsymbol{\delta}-m_{\delta}^{2}\boldsymbol{\delta}^{2}) \\
                &-\frac{1}{4}W^{\mu\nu}W_{\mu\nu}+\frac{1}{2}m_{\omega}^{2}\omega^{\mu}\omega_{\mu}
                 -\frac{1}{4}\boldsymbol{R}^{\mu\nu}\boldsymbol{R}_{\mu\nu}+\frac{1}{2}m_{\rho}^{2}\boldsymbol{\rho}^{\mu}\boldsymbol{\rho}_{\mu}
                 -\frac{1}{4}\Phi^{\mu\nu}\Phi_{\mu\nu}+\frac{1}{2}m_{\phi}^{2}\phi^{\mu}\phi_{\mu} \\
                &+\sum_{l}\bar{\psi}_{l}(i\gamma_{\mu}\partial^{\mu}-m_{l})\psi_{l},
\end{aligned}
\end{equation}
where~$\boldsymbol{\tau}_{B}$~is the Pauli matrices for isospin of the baryon species~$B$, and~$M_{B}$~and~$m_{l}$~represent the baryon and lepton masses, respectively. $\psi_{B(l)}$~is the Dirac field of the baryon species~$B$~or the lepton species~$l$. $\sigma$, $\omega_{\mu}$, $\boldsymbol{\rho}_{\mu}$, $\boldsymbol{\delta}$, and~$\phi_{\mu}$~denote the quantum fields of mesons. The antisymmetric field strength tensors~($W_{\mu\nu}$, $\boldsymbol{R}_{\mu\nu}$, and $\Phi_{\mu\nu}$)~of vector mesons~($\omega$, $\rho$, and~$\phi$)~are
\begin{equation}\label{equ:meson_fieldtensor}
\begin{aligned}
    W_{\mu\nu}       &= \partial_{\mu}\omega_{\nu}-\partial_{\nu}\omega_{\mu}, \\
    \boldsymbol{R}_{\mu\nu} &= \partial_{\mu}\vec{\rho}_{\nu}-\partial_{\nu}\vec{\rho}_{\mu}, \\
    \Phi_{\mu\nu}    &= \partial_{\mu}\phi_{\nu}-\partial_{\nu}\phi_{\mu}.
\end{aligned}
\end{equation}

Under the mean-field approximation, all quantum fluctuations of meson fields are neglected and the meson fields are treated as classical fields. Then, the equations of motion of various mesons are obtained via the Euler-Lagrange equation
\begin{equation}\label{equ:equ_motion_meson}
\begin{aligned}
    m_{\sigma}^{2}\sigma &= -\sum_{B}\Gamma_{\sigma B}(\rho_{B})\rho_{s}^{B}, \\
    m_{\omega}^{2}\omega &= \sum_{B}\Gamma_{\omega B}(\rho_{B})\rho_{v}^{B}, \\
    m_{\rho}^{2}\rho     &= \sum_{B}\Gamma_{\rho B}(\rho_{B})\rho_{v}^{B}\tau_{B}^{3}, \\
    m_{\delta}^{2}\delta &= -\sum_{B}\Gamma_{\delta B}(\rho_{B})\rho_{s}^{B}\tau_{B}^{3}, \\
    m_{\phi}^{2}\phi     &= \sum_{B}\Gamma_{\phi B}(\rho_{B})\rho_{v}^{B},
\end{aligned}
\end{equation}
where $\tau_{B}^{3}$ is the isospin projection of the baryon species $B$. The vector density $\rho_{v}^{B}$ and scalar density $\rho_{s}^{B}$ of the baryon species $B$ read
\begin{equation}\label{equ:scalar+vector_density}
\begin{aligned}
    \rho_{v}^{B} &= \frac{1}{\pi^{2}}\int_{0}^{k_{\mathrm{F}}^{B}}k^{2}\mathrm{d}k = \frac{(k_{\mathrm{F}}^{B})^{3}}{3\pi^{2}}, \\
    \rho_{s}^{B} &= \frac{M_{B}^{\star}}{\pi^{2}}\int_{0}^{k_{\mathrm{F}}^{B}}\frac{k^{2}\mathrm{d}k}{\sqrt{k^{2}+M_{B}^{\star 2}}}
                   = \frac{(M_{B}^{\star})^{3}}{2\pi^{2}}\left[ q\sqrt{1+q^{2}}-\ln(q+\sqrt{1+q^{2}}) \right],
                     \quad q=\frac{k_{\mathrm{F}}^{B}}{M_{B}^{\star}},
\end{aligned}
\end{equation}
with the Fermi momentum~$k_{\mathrm{F}}^{B}$~and effective mass~$M_{B}^{\star}=M_{B}+\Gamma_{\sigma B}(\rho_{B})\sigma+\Gamma_{\delta B}(\rho_{B})\delta\tau_{B}^{3}$~of the baryon species~$B$.

Inside neutron star cores, the baryon and lepton compositions satisfy the~$\beta$-equilibrium conditions
\begin{equation}\label{equ:beta_equilibrium}
\begin{aligned}
    &\mu_{n}-q_{B}\mu_{e} = \mu_{B}, \\
    &\mu_{\mu} = \mu_{e},
\end{aligned}
\end{equation}
where~$q_{B}$~is the charge of the baryon species~$B$. $\mu_{B}$~($\mu_{l}$), which stands for the chemical potential of the baryon species~$B$~(the lepton species~$l$), is given by
\begin{equation}\label{equ:DDRMF_chem_potential}
\begin{aligned}
    \mu_{B}(k_{\mathrm{F}}^{B}) &= \sqrt{(k_{\mathrm{F}}^{B})^{2}+M_{B}^{\star 2}}
                                  +\Gamma_{\omega B}(\rho_{B})\omega+\Gamma_{\rho B}(\rho_{B})\rho\tau_{3}^{B}
                                  +\Gamma_{\phi B}(\rho_{B})\phi+\Sigma_{R}, \\
    \mu_{l}(k_{\mathrm{F}}^{l}) &= \sqrt{(k_{\mathrm{F}}^{l})^{2}+M_{l}^{\star 2}}.
\end{aligned}
\end{equation}
The density dependence of coupling constants between baryons and mesons leads to the rearrangement term~$\Sigma_{R}$~in Eq.~(\ref{equ:DDRMF_chem_potential})~\citep{Fuchs1995_PRC52-3043},
\begin{equation}\label{equ:rearr_term}
    \Sigma_{R} = \sum_{B}\left[ -\frac{\partial\Gamma_{\sigma B}(\rho_{B})}{\partial\rho_{B}}\sigma\rho_{s}^{B}
                     -\frac{\partial\Gamma_{\omega B}(\rho_{B})}{\partial\rho_{B}}\omega\rho_{v}^{B}
                     -\frac{\partial\Gamma_{\rho B}(\rho_{B})}{\partial\rho_{B}}\rho\rho_{v}^{B}\tau_{3}^{B}
                     -\frac{\partial\Gamma_{\delta B}(\rho_{B})}{\partial\rho_{B}}\delta\rho_{s}^{B}\tau_{3}^{B}
                     -\frac{\partial\Gamma_{\phi B}(\rho_{B})}{\partial\rho_{B}}\phi\rho_{v}^{B}\right].
\end{equation}

Several different forms have been proposed for the density dependence of coupling constants~\citep{deJong1998_PRC57-3099,Typel1999_NPA656-331}. For all effective interactions we use in this work, the coupling constants between baryons and isoscalar mesons ($\sigma$ and $\omega$) are given by
\begin{equation}\label{equ:DD_form1}
    \Gamma_{iB}(x)=\Gamma_{iB}(\rho_{0})a_{i}\frac{1+b_{i}(x+d_{i})^{2}}{1+c_{i}(x+e_{i})^{2}},\quad x=\rho_{B}/\rho_{0}. \\
\end{equation}
For the coupling constants between baryons and isovector mesons~($\rho$~and~$\delta$), the density dependence in DD-ME$\delta$~takes the form of Eq.~(\ref{equ:DD_form1})~but in other effective interactions reads
\begin{equation}\label{equ:DD_form2}
    \Gamma_{iB}(x)=\Gamma_{iB}(\rho_{0})\mathrm{exp}^{-a_{i}(x-1)},\quad x=\rho_{B}/\rho_{0}.
\end{equation}

For the coupling constants between hyperons and mesons, we apply the relations from the SU(6) naive quark model for~$\omega$, $\rho$, $\delta$, and~$\phi$~\citep{Shao2010_PRC82-055801,Schaffner1996_PRC53-1416}
\begin{equation}\label{equ:Ym_coup_const}
\begin{aligned}
    & \Gamma_{\omega\Lambda} = \Gamma_{\omega\Sigma} = 2\Gamma_{\omega\Xi} = \frac{2}{3}\Gamma_{\omega N}, \\
    & \Gamma_{\rho\Lambda} = 0,\quad \Gamma_{\rho\Sigma} = 2\Gamma_{\omega\Xi} = 2\Gamma_{\omega N}, \\
    & \Gamma_{\delta\Lambda} = 0,\quad \Gamma_{\delta\Sigma} = 2\Gamma_{\delta\Xi} = 2\Gamma_{\delta N}, \\
    & \Gamma_{\phi N} = 0,\quad 2\Gamma_{\phi\Lambda} = 2\Gamma_{\phi\Sigma} = \Gamma_{\phi\Xi} = -\frac{2\sqrt{2}}{3}\Gamma_{\omega N}.
\end{aligned}
\end{equation}
As for the~$\sigma$-meson, we determine the coupling constants between hyperons and $\sigma$ by fitting empirical hypernuclear potentials using the following formula
\begin{equation}\label{equ:Ys_coup_const}
   U_{Y}^{(N)} = R_{\sigma Y}\Gamma_{\sigma N}^{0}\sigma^{0}+R_{\omega Y}\Gamma_{\omega N}^{0}\omega^{0}+\Sigma_{R}^{0},
\end{equation}
where~$\Gamma_{\sigma N}^{0}$, $\Gamma_{\omega N}^{0}$, $\sigma^{0}$, $\omega^{0}$, and~$\Sigma_{R}^{0}$~are the values of symmetric nuclear matter at the saturation density and $R_{\sigma Y}$ ($R_{\omega Y}$) is the ratio of~$\Gamma_{\sigma Y}$~to~$\Gamma_{\sigma N}$~($\Gamma_{\omega Y}$~to~$\Gamma_{\omega N}$). We choose~$U_{\Lambda}^{(N)}=-30$ MeV \citep{Schaffner-Bielich2000_PRC62-034311,Wang2010_PRC81-025801}, $U_{\Sigma}^{(N)}=+30$ MeV, and $U_{\Xi}^{(N)}=-15$ MeV \citep{Ishizuka2008_JPG35-085201,Wang2010_PRC81-025801}~ in the present work.

The charge neutrality condition~$\rho_{v}^{p}+\rho_{v}^{\Sigma^{+}}=\rho_{v}^{e}+\rho_{v}^{\mu}+\rho_{v}^{\Sigma^{-}}+\rho_{v}^{\Xi^{-}}$~and baryon number conservation condition~$\rho_{B}=\sum_{B}\rho_{v}^{B}$~are satisfied inside neutron star cores. With these two conditions, we can obtain the meson fields and Fermi momenta of baryons and leptons by solving the non-linear coupled equations (\ref{equ:equ_motion_meson}) and ~(\ref{equ:beta_equilibrium}) self-consistently at a given baryon number density. Furthermore, the energy density $\varepsilon$ and pressure $P$ can be calculated by using the energy--momentum tensor,
\begin{equation}\label{equ:DDRMF_EoS}
\begin{aligned}
    \varepsilon &= \frac{1}{2}m_{\sigma}^{2}\sigma^{2}+\frac{1}{2}m_{\omega}^{2}\omega^{2}+\frac{1}{2}m_{\rho}^{2}\rho^{2}
                    +\frac{1}{2}m_{\delta}^{2}\delta^{2}+\frac{1}{2}m_{\phi}^{2}\phi^{2}
                    +\sum_{B}\varepsilon_{\mathrm{kin}}^{B}+\sum_{l}\varepsilon_{\mathrm{kin}}^{l}, \\
        P       &= -\frac{1}{2}m_{\sigma}^{2}\sigma^{2}+\frac{1}{2}m_{\omega}^{2}\omega^{2}+\frac{1}{2}m_{\rho}^{2}\rho^{2}
                    -\frac{1}{2}m_{\delta}^{2}\delta^{2}+\frac{1}{2}m_{\phi}^{2}\phi^{2}+\rho\Sigma_{N}^{r}
                    +\sum_{B}P_{\mathrm{kin}}^{B}+\sum_{l}P_{\mathrm{kin}}^{l},
\end{aligned}
\end{equation}
where~$\varepsilon_{\mathrm{kin}}^{B}$~($\varepsilon_{\mathrm{kin}}^{l}$)~and~$P_{\mathrm{kin}}^{B}$~($P_{\mathrm{kin}}^{l}$)~are the contributions from kinetic energy,
\begin{equation}\label{equ:DDRMF_EoS_kin}
\begin{aligned}
    \varepsilon_{\mathrm{kin}} &= \frac{k_{\mathrm{F}}^{4}}{\pi^{2}}\left[ \left(1+\frac{z^{2}}{2}\right)\frac{\sqrt{1+z^{2}}}{4}
                                   -\frac{z^{4}}{8}\ln\left(\frac{1+\sqrt{1+z^{2}}}{z}\right) \right], \\
          P_{\mathrm{kin}}     &= \frac{k_{\mathrm{F}}^{4}}{3\pi^{2}}\left[ \left(1-\frac{3z^{2}}{2}\right)\frac{\sqrt{1+z^{2}}}{4}
                                   +\frac{3z^{4}}{8}\ln\left(\frac{1+\sqrt{1+z^{2}}}{z}\right) \right], \quad z=\frac{1}{q}.
\end{aligned}
\end{equation}
Usually, the EoS is the pressure as a function of the energy density. The squared sound velocity~$v_{s}^{2}$~is related to EoS by
\begin{equation}\label{equ:DDRMF_soundvelocity}
    v_{s}^{2} = \frac{\partial P}{\partial\varepsilon}.
\end{equation}

The mass--radius~($M$--$R$)~relation of a stationary neutron star is obtained by solving the Tolman--Oppenheimer--Volkoff (TOV) equation~\citep{Tolman1939_PR055-364,Oppenheimer1939_PR055-374}
\begin{equation}\label{equ:TOV}
\begin{aligned}
    \frac{\mathrm{d}P}{\mathrm{d}r} &= -\frac{\left[ P(r)+\varepsilon(r) \right]\left[ M(r)+4\pi r^{3}P(r) \right]}{r\left[ r-2M(r) \right]}, \\
    \frac{\mathrm{d}M}{\mathrm{d}r} &= 4\pi r^{2}\varepsilon(r),
\end{aligned}
\end{equation}
with a given EoS as input, where~$r$~is the distance from the center. Given a central density~$\rho_{c}$~at~$r=0$, the TOV equation is integrated from~$r=0$~to~$r=R$~where the pressure is zero. $R$~is defined as the radius of the neutron star and~$M(R)$~is the gravitational mass.

\section{Results and Discussions} \label{sec:resul}
\subsection{DDRMF effective interactions} \label{subsec:parameter_sets}
In this work, the neutron star properties are calculated by using eight DDRMF effective interactions under different assumptions on the interior composition. The $\delta$-meson is taken into account in DD-ME$\delta$~and DDVTD. The tensor coupling is included in DDVT and DDVTD, but we do not consider its contribution because the tensor coupling effects vanish in nuclear matter~\citep{Typel2020_EPJA56-160}. In Table~\ref{tab:DDRMF_models}, the saturation properties of symmetric nuclear matter calculated with different effective interactions are listed, including the saturation density~$\rho_{0}$, binding energy per particle~$E/A$, incompressibility~$K_{0}$, symmetry energy~$E_{\mathrm{sym}}$, slope of symmetry energy~$L$, and effective mass of neutron~$M_{n}^{\star}/M_{n}$.
\begin{deluxetable*}{lllcccc}
\tablenum{1}
\tablecaption{Saturation properties of nuclear matter for different DDRMF effective interactions.\label{tab:DDRMF_models}}
\tablewidth{0pt}
\tablehead{
\colhead{Effective Interaction} & \colhead{$\rho_{0}$} & \colhead{$E/A$} & \colhead{$K_{0}$} & \colhead{$E_{\mathrm{sym}}$} & \colhead{$L$}
& \colhead{$M_{n}^{\star}/M_{n}$} \\
\colhead{}          & \colhead{(fm$^{-3}$)}& \colhead{(MeV)} & \colhead{(MeV)}   & \colhead{(MeV)}              & \colhead{(MeV)}
& \colhead{}
}
\decimalcolnumbers
\startdata
DD-ME2        & 0.152 & $-$16.14  & 251.1 & 32.30 &51.26 & 0.572 \\
DD-ME$\delta$ & 0.152 & $-$16.12  & 219.1 & 32.35 &52.85 & 0.609 \\
PKDD          & 0.150 & $-$16.27  & 262.2 & 36.86 &90.21 & 0.570 \\
TW99          & 0.153 & $-$16.25  & 240.2 & 32.77 &55.31 & 0.555 \\
DD-MEX        & 0.152 & $-$16.14  & 267.1 & 32.27 &49.69 & 0.556 \\
DDV           & 0.1511& $-$16.097 & 239.5 & 33.59 &69.65 & 0.586 \\
DDVT          & 0.1536& $-$16.924 & 240.0 & 31.56 &42.35 & 0.667 \\
DDVTD         & 0.1536& $-$16.915 & 239.9 & 31.82 &42.58 & 0.667 \\
\enddata
\tablecomments{The saturation properties we list here include the saturation density~$\rho_{0}$~(fm$^{-3}$), binding energy per particle $E/A$ (MeV), incompressibility $K_{0}$ (MeV), symmetry energy $E_{\mathrm{sym}}$ (MeV), slope of symmetry energy $L$ (MeV), and effective mass of neutron $M_{n}^{\star}/M_{n}$.}
\end{deluxetable*}

The incompressibility~$K_{0}$~and slope of symmetry energy~$L$~significantly affect the EoS and macroscopic properties (e.g., mass and radius) of neutron stars~\citep{Chen2014_PRC90-044305,Biswal2019_ApJ885-25,Ji2019_PRC100-045801,Choi2021_ApJ909-156}. Several constraints on~$K_{0}$~and~$L$~were obtained from terrestrial experiments and astrophysical observations. We first check whether these effective interactions can be ruled out by these constraints. Among the eight effective interactions, the values of $K_{0}$ range from 219.1 MeV (DD-ME$\delta$) to 267.1 MeV(DD-MEX), which satisfy the recent constraint, $215~\mathrm{MeV}\leqslant K_{0}\leqslant260$ MeV, from \citet{Choi2021_ApJ909-156} except for PKDD and DD-ME2. For $L$, the values range from 42.35 MeV (DDVT) to 90.21 MeV (PKDD), which fulfill the constraint in \citet{Oertel2017_RMP89-015007} ($L=58.7\pm28.1$ MeV)  and \citet{Choi2021_ApJ909-156} (40 MeV$\leqslant L \leqslant$ 85 MeV) except for PKDD. Besides, in \citet{Yan2019_RAA19-072}, the observations of glitching pulsars were used to constrain the symmetry energy and incompressibility and it was found that the lower limits of $K_{0}$ and $L$ are 215 MeV and 67 MeV, respectively; only PKDD and DDV can meet the constraint of $L$. Reed~\emph{et al}. obtained a value of $L=$ 106 $\pm$ 37 MeV by analyzing the neutron skin thickness of $^{208}$Pb \citep{Reed2021_PRL126-172503}, which is consistent with the lower limit of $L$ in~\citet{Yan2019_RAA19-072}. Due to the uncertainty of these constraints, it is difficult to judge which effective interaction is the best one. Investigating the universal effects of $\phi$-meson on hyperon stars by using these effective interactions with large uncertainties of $K_{0}$ and $L$ is necessary and meaningful.
\begin{deluxetable*}{lrrr}
\tablenum{2}
\tablecaption{Coupling constants between hyperons and $\sigma$ for different DDRMF effective interactions.\label{tab:coup_const_Ym}}
\tablewidth{0pt}
\tablehead{
\colhead{Effective Interaction} & \colhead{$R_{\sigma\Lambda}$} & \colhead{$R_{\sigma\Sigma}$} & \colhead{$R_{\sigma\Xi}$}
}
\decimalcolnumbers
\startdata
DD-ME2        & $0.620035$ & $0.470799$ & $0.315064$\\
DD-ME$\delta$ & $0.625095$ & $0.461759$ & $0.323150$\\
PKDD          & $0.620933$ & $0.472392$ & $0.315603$\\
TW99          & $0.617016$ & $0.473473$ & $0.311023$\\
DD-MEX        & $0.617628$ & $0.474792$ & $0.311198$\\
DDV           & $0.622105$ & $0.467624$ & $0.318478$\\
DDVT          & $0.631152$ & $0.439371$ & $0.335579$\\
DDVTD         & $0.631716$ & $0.439687$ & $0.335942$\\
\enddata
\end{deluxetable*}
\begin{deluxetable*}{lllrlrlrlr}
\tablenum{3}
\tablecaption{Hyperon thresholds calculated with different DDRMF effective interactions for $npe\mu Y$ and $npe\mu Y\phi$ matter. \label{tab:onset_hyperon}}
\tablewidth{0pt}
\tablehead{
\colhead{Matter} & \colhead{Effective Interaction} & \multicolumn2c{1$^{\mathrm{st}}~Y$} & \multicolumn2c{2$^{\mathrm{nd}}~Y$} & \multicolumn2c{3$^{\mathrm{rd}}~Y$} & \multicolumn2c{4$^{\mathrm{th}}~Y$} \\
\colhead{      } & \colhead{         } & \colhead{$Y$} & \colhead{$\rho_{\mathrm{Thold}}$} & \colhead{$Y$} & \colhead{$\rho_{\mathrm{Thold}}$} & \colhead{$Y$} & \colhead{$\rho_{\mathrm{Thold}}$} & \colhead{$Y$} & \colhead{$\rho_{\mathrm{Thold}}$} \\
\cmidrule(r){3-4} \cmidrule(r){5-6} \cmidrule(r){7-8} \cmidrule(r){9-10}
\colhead{      } & \colhead{         } & \colhead{   } & \colhead{(fm$^{-3}$)} & \colhead{   } & \colhead{(fm$^{-3}$)} & \colhead{   } & \colhead{(fm$^{-3}$)} & \colhead{   } & \colhead{(fm$^{-3}$)}
}
\decimalcolnumbers
\startdata
\multirow{8}*{$npe\mu Y$} &
 DD-ME2        & $\Lambda$     & $0.3314$  & $\Xi^{-}$     & $0.3684$  & $\Sigma^{-}$ & $0.3756$   & $\Xi^{0}$   & 0.7847\\
 & DD-ME$\delta$ & $\Sigma^{-}$  & $0.3749$  & $\Lambda$     & $0.3782$  & $\Xi^{-}$    & $0.4862$   & $\Xi^{0}$   & 1.3355\\
 & PKDD          & $\Lambda$     & $0.3159$  & $\Xi^{-}$     & $0.3799$  & $\Xi^{0}$    & $0.8272$   &             & \\
 & TW99          & $\Lambda$     & $0.3631$  & $\Sigma^{-}$  & $0.3907$  & $\Xi^{-}$    & $0.4369$   & $\Xi^{0}$   & 1.0359\\
 & DD-MEX        & $\Lambda$     & $0.3212$  & $\Sigma^{-}$  & $0.3540$  & $\Xi^{-}$    & $0.3552$   & $\Xi^{0}$   & 0.7492\\
 & DDV           & $\Lambda$     & $0.3386$  & $\Xi^{-}$     & $0.3985$  & $\Sigma^{-}$ & $0.4286$   & $\Xi^{0}$   & 1.0737\\
 & DDVT          & $\Lambda$     & $0.4020$  & $\Xi^{-}$     & $0.4539$  & $\Xi^{0}$    & $1.1343$   &             & \\
 & DDVTD         & $\Lambda$     & $0.3998$  & $\Xi^{-}$     & $0.4540$  & $\Xi^{0}$    & $1.1792$   &             & \\
\hline
\multirow{8}*{$npe\mu Y\phi$} &
 DD-ME2        & $\Lambda$     & $0.3314$  & $\Sigma^{-}$  & $0.3745$  & $\Xi^{-}$    & $0.3764$   & $\Xi^{0}$   & 1.1580\\
 & DD-ME$\delta$ & $\Sigma^{-}$  & $0.3749$  & $\Lambda$     & $0.3785$  & $\Xi^{-}$    & $0.5448$   &             & \\
 & PKDD          & $\Lambda$     & $0.3159$  & $\Xi^{-}$     & $0.3957$  & $\Sigma^{-}$ & $0.4977$   & $\Xi^{0}$   & 1.2315\\
 & TW99          & $\Lambda$     & $0.3631$  & $\Sigma^{-}$  & $0.3925$  & $\Xi^{-}$    & $0.4733$   & $\Xi^{0}$   & 1.4821\\
 & DD-MEX        & $\Lambda$     & $0.3212$  & $\Sigma^{-}$  & $0.3569$  & $\Xi^{-}$    & $0.3653$   & $\Xi^{0}$   & 1.1108\\
 & DDV           & $\Lambda$     & $0.3386$  & $\Xi^{-}$     & $0.4119$  & $\Sigma^{-}$ & $0.4170$   & $\Xi^{0}$   & 1.5840\\
 & DDVT          & $\Lambda$     & $0.4020$  & $\Xi^{-}$     & $0.4623$  & $\Sigma^{-}$ & $0.5186$   & $\Xi^{0}$   & 1.6452\\
 & DDVTD         & $\Lambda$     & $0.3998$  & $\Xi^{-}$     & $0.4629$  & $\Sigma^{-}$ & $0.5189$   & $\Xi^{0}$   & 1.7230\\
\enddata
\tablecomments{$\rho_{\mathrm{Thold}}$ is the hyperon threshold density.}
\end{deluxetable*}
\subsection{Effects of $\phi$-meson on the interior composition} \label{subsec:phi_on_interior_composition}
\begin{figure}[ht!]
\centering
\includegraphics[scale=0.5]{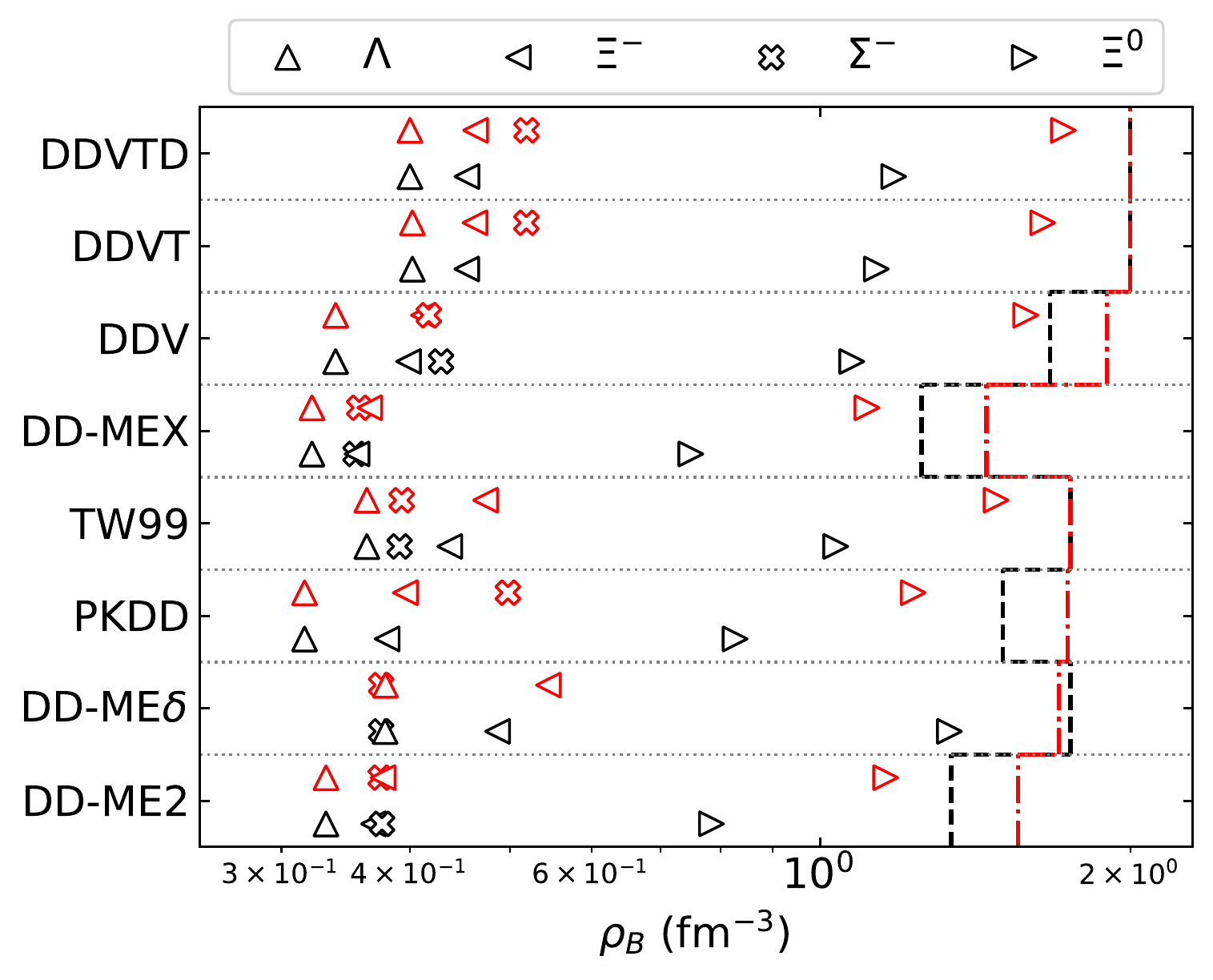}
\caption{Hyperon thresholds of $npe\mu Y$ (black symbol) and $npe\mu Y\phi$ (red symbol)~matters calculated with different effective interactions. The dash lines stand for truncation densities before which the maximum masses have been reached. \label{fig:NS_hyper_threshold}}
\end{figure}

In Table~\ref{tab:coup_const_Ym}, we list the ratio~$R_{\sigma Y}$~for different effective interactions. Combining with the hyperon--meson coupling constants given in Table~\ref{tab:coup_const_Ym} and Eq.~(\ref{equ:Ym_coup_const}), within a given density range, the baryon and lepton fractions as a function of~$\rho_{B}$~for various effective interactions are obtained by self-consistently solving the non-linear coupled equations consisting of~Eq.~(\ref{equ:equ_motion_meson}), Eq.~(\ref{equ:beta_equilibrium}), the charge neutrality condition, and the baryon number conservation condition. The hyperon thresholds can be extracted easily from the baryon fractions. The hyperon thresholds of~$npe\mu Y$~and~$npe\mu Y\phi$~matters calculated with different effective interactions are listed in Table~\ref{tab:onset_hyperon}~and shown in Fig.~\ref{fig:NS_hyper_threshold}.

For the $npe\mu Y$ matter, within the density range we consider, it is noticed that not all hyperons appear as density increases and the orders and thresholds of the appearance of various hyperons with different effective interactions are very different. In our calculations, only~$\Lambda$, $\Sigma^{-}$, $\Xi^{0}$, and~$\Xi^{-}$~are likely to be populated. In general, both effective interaction and hyperon properties influence the hyperon thresholds in different manners. This can be understood by the threshold equation
\begin{equation}\label{equ:threshold_equ}
  \mu_{n}-q_{B}\mu_{e} \geqslant M_{B}+\Gamma_{\sigma B}\sigma+\Gamma_{\delta B}\delta\tau_{3}^{B}
                                 +\Gamma_{\omega B}\omega+\Gamma_{\rho B}\rho\tau_{3}^{B}.
\end{equation}
Once the condition in Eq.~(\ref{equ:threshold_equ}) is fulfilled, the hyperon is populated. A hyperon with smaller mass is more mass-favored, according to the first term on the right hand side. Negatively charged hyperons are charge-favored because they can replace the role of the neutral baryons and leptons at the top of the Fermi sea~\citep{Glendenning1985_ApJ293-470}. A hyperon having the opposite (same) sign as~$\tau_{3}$~of the neutron is isospin-favored when the sign of~$\Gamma_{\delta B}\delta+\Gamma_{\rho B}\rho$~is fixed as negative (positive). The hypernuclear potential affects the thresholds by changing $R_{\sigma Y}$ through the relation in Eq.~(\ref{equ:Ys_coup_const}). The effective interaction mainly determines the behaviors of nucleonic matter before the first hyperon is populated, and further affects the hyperon thresholds together with hyperon properties.

\begin{figure}[ht!]
\centering
\includegraphics[scale=0.5]{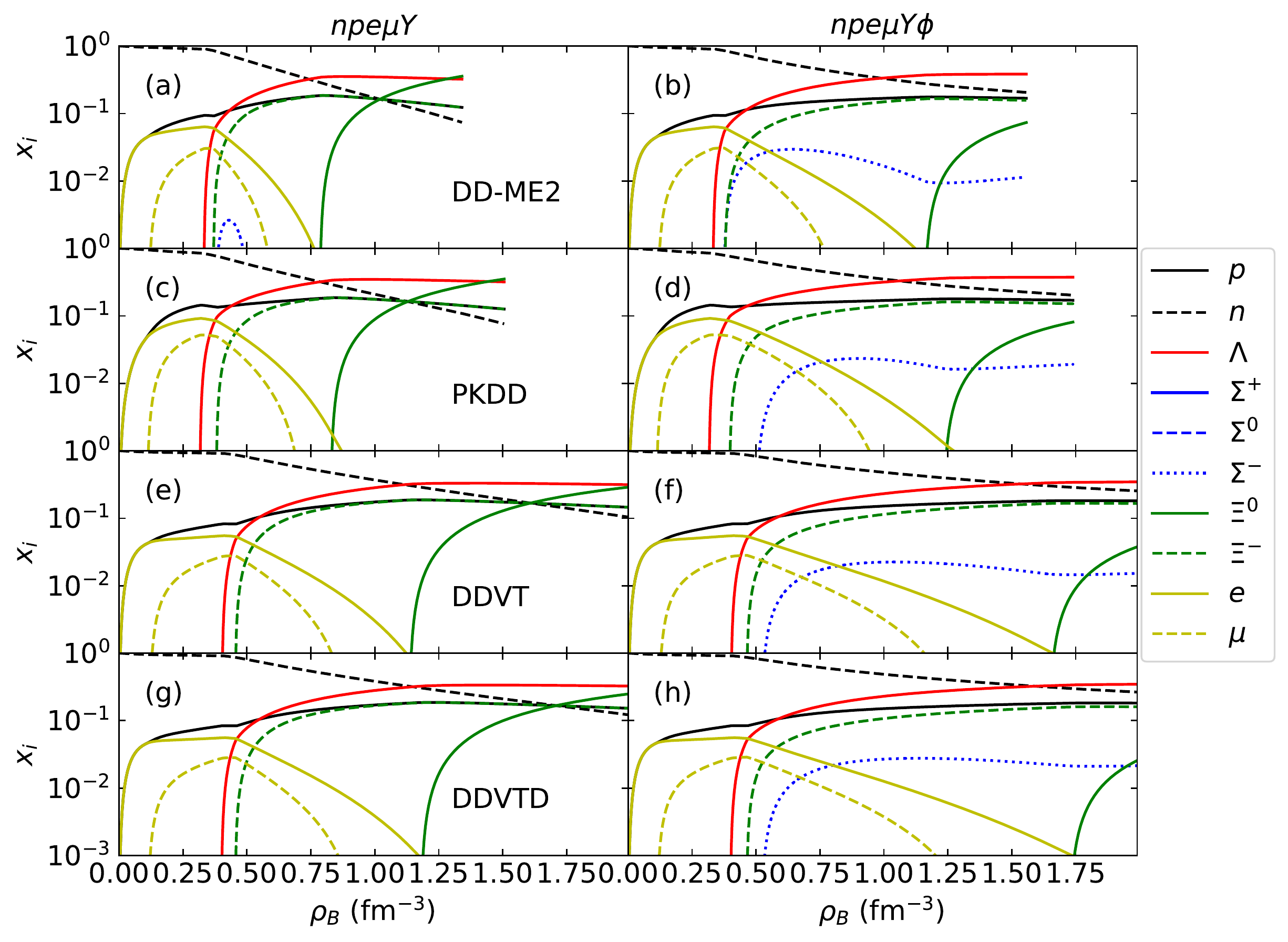}
\caption{Particle fractions of baryons as a function of baryon number density with DD-ME2, PKDD, DDVT, and DDVTD. Panels (a), (c), (e), and (g) are for the~$npe\mu Y$~matter and (b), (d), (f), and (h) are for the~$npe\mu Y\phi$~matter. The two panels on the same row represent the results with the same effective interaction. The particle fractions of leptons, defined as $\rho_{l}/\rho_{B}$, as a function of the baryon number density are also plotted as yellow solid line (electron) and yellow dotted line (muon). \label{fig:baryon_fraction}}
\end{figure}

Now we discuss the effects of $\phi$-meson on hyperon thresholds in the $npe\mu Y\phi$ matter. In Fig.~\ref{fig:baryon_fraction}, we show the particle fractions ($X_{i}=\rho_{v}^{i}/\rho_{B}, i=n, p, \Lambda, \Sigma^{\pm, 0}$, and $\Xi^{0,-}$) with DD-ME2, PKDD, DDVT, and DDVTD. The main differences of hyperon thresholds between $npe\mu Y$ and $npe\mu Y\phi$ matters can be found in Fig.~\ref{fig:NS_hyper_threshold}, Fig.~\ref{fig:baryon_fraction}, and Table~\ref{tab:onset_hyperon}: (a) Compared with the $npe\mu Y$ matter, the hyperon thresholds of the $npe\mu Y\phi$ matter are shifted to higher densities (especially for $\Xi$) except for the first appearing hyperon; (b) the order of the thresholds of $\Sigma^{-}$ and $\Xi^{-}$ is reversed after the $\phi$-meson is included in calculations with DD-ME2; (c) $\Sigma^{-}$~appears in the $npe\mu Y\phi$ matter but not in the $npe\mu Y$ matter with PKDD, DDVT, and DDVTD.
\begin{figure}[ht!]
\centering
\includegraphics[scale=0.5]{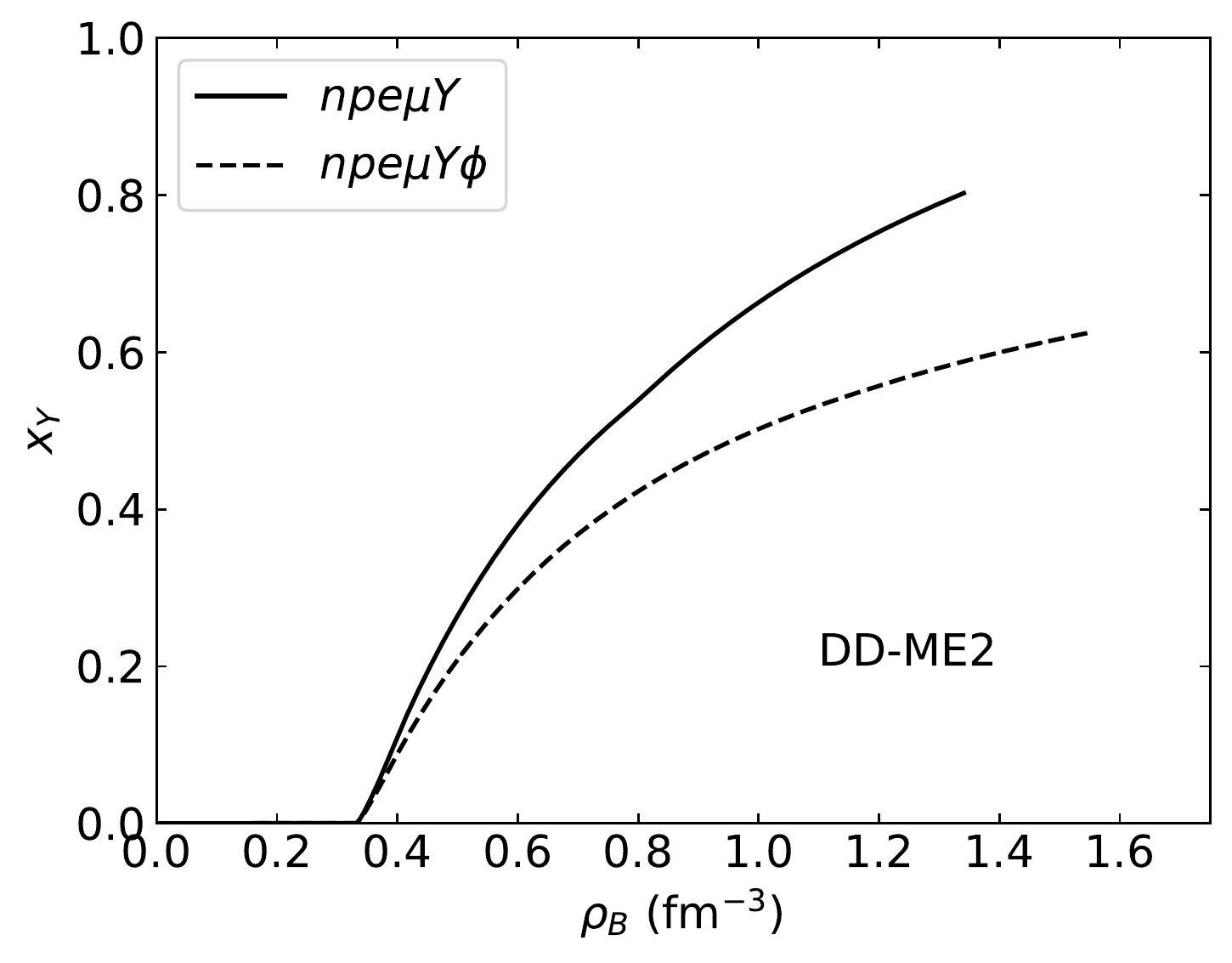}
\caption{Total hyperon fraction as a function of baryon number density with DD-ME2 in~$npe\mu Y$~and~$npe\mu Y\phi$~matters.\label{fig:Yfrac_DD-ME2}}
\end{figure}

In order to understand the effects of $\phi$-meson, we display the total hyperon fraction ($X_{Y}=\sum_{i}\rho_{v}^{i}/\rho_{B}, i=\Lambda, \Sigma^{\pm, 0}$, and $\Xi^{0,-}$) as a function of baryon number density in Fig.~\ref{fig:Yfrac_DD-ME2}. Taking DD-ME2 as an example, $X_{Y}$ is reduced significantly when $\phi$-meson is considered in comparison with the case of the $npe\mu Y$ matter. The~$\phi$-meson enhances the repulsive interaction between hyperons and increases the energy of hypernuclear matter, causing the reduction of the particle fractions of hyperons that are energy-favored; note that this is also found in the NLRMF model~\citep{Banik2014_ApJSupp214-22}. The~$\beta$-equilibrium conditions of hyperons in Eq.~(\ref{equ:beta_equilibrium})~can be also used to explain the reduction of $X_{Y}$ in the $npe\mu Y\phi$ matter. Compared to the $npe\mu Y$ matter, a positive term is added on the right hand side of the $\beta$-equilibrium condition when $\phi$-meson is included. To ensure that the $\beta$-equilibrium condition is fulfilled, the Fermi momentum of hyperon needs to become smaller, which leads to the fact that $X_{Y}$ is reduced. The results and discussions of other effective interactions are similar to DD-ME2.

In Fig.~\ref{fig:NS_hyper_threshold}~and Fig.~\ref{fig:baryon_fraction}, we can see that hyperon thresholds are shifted to higher densities for all effective interactions when $\phi$-meson is considered, except for the first appearing hyperon. The $\phi$-meson mediates the repulsive interaction between hyperons and takes effect only after the first appearing hyperon is populated. Therefore the first hyperon threshold remains unchanged strictly. For the $npe\mu Y\phi$ matter, the threshold equation of the hyperon species~$B$~is written as
\begin{equation}\label{equ:threshold_equ_phi}
  \mu_{n}-q_{B}\mu_{e} \geqslant M_{B}+\Gamma_{\sigma B}\sigma+\Gamma_{\delta B}\delta\tau_{3}^{B}
                                 +\Gamma_{\omega B}\omega+\Gamma_{\rho B}\rho\tau_{3}^{B}+\Gamma_{\phi B}\phi.
\end{equation}
The $\phi$-meson contributes a positive term to the right hand side of Eq.~(\ref{equ:threshold_equ}). A larger $\mu_{n}$ at higher density is required due to the larger value of the right hand side in Eq.~(\ref{equ:threshold_equ_phi}), causing that the hyperon threshold is shifted to a higher density.

The $\phi$-meson has a greater effect on the threshold density of $\Xi$ because its strangeness number is $-2$ and the coupling strength between $\Xi$ and $\phi$ is twice of those for other hyperons~\citep{Providencia2019_FASS6-13}. This can be used to explain the reversal of the order of hyperon appearance with DD-ME2. The threshold densities of $\Xi$ increase more significantly than those of other hyperons because the contribution of $\phi$-meson in Eq.~(\ref{equ:threshold_equ_phi}) for $\Xi$ is twice as much as those of other hyperons. The reason why the order of the appearance of $\Sigma^{-}$ and $\Xi^{-}$ is reversed in the calculations of the DD-ME2 is that the $\phi$-meson has a greater impact on $\Xi^{-}$ so that the threshold density of $\Xi^{-}$~becomes larger than that of $\Sigma^{-}$.
\begin{figure}[ht!]
\centering
\includegraphics[scale=0.5]{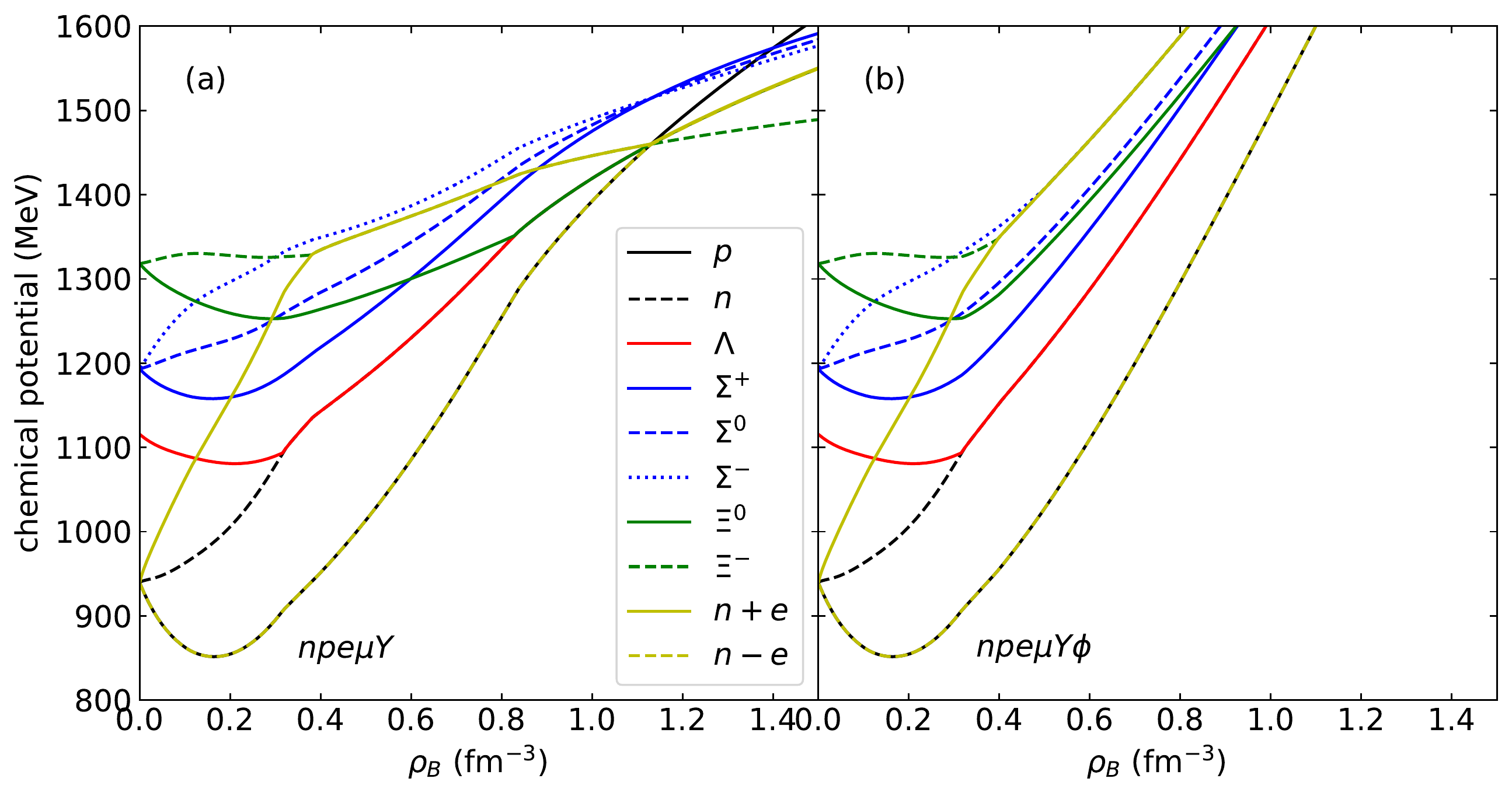}
\caption{Chemical potential of baryons as a function of baryon number density with PKDD in $npe\mu Y$ and $npe\mu Y\phi$ matters. The labels ``$n+e$'' and ``$n-e$'' represent the two cases of the left hand side in the first equation of Eq.~(\ref{equ:beta_equilibrium}),  $\mu_{n}+\mu_{e}$ and $\mu_{n}-\mu_{e}$.\label{fig:chem_PKDD}}
\end{figure}

The $\phi$-meson may change the hyperon species inside neutron stars. To explore the reason why $\Sigma^{-}$ appears in the $npe\mu Y\phi$ matter but not in the $npe\mu Y$ matter with PKDD, DDVT, and DDVTD, we show the chemical potentials of baryons as a function of the baryon number density in Fig.~\ref{fig:chem_PKDD}, taking PKDD as an example. In Fig.~\ref{fig:chem_PKDD}, the intersection of the chemical potential of a hyperon and $\mu_{n}+\mu_{e}$ or $\mu_{n}-\mu_{e}$ mean that the corresponding hyperon threshold is reached, and the two chemical potential curves which keep coincident above the threshold density ensure that the $\beta$-equilibrium condition is satisfied accordingly. For $\Sigma^{-}$ and $\Xi^{-}$, their $\beta$-equilibrium conditions are $\mu_{\Sigma^{-}}=\mu_{\Xi^{-}}=\mu_{n}+\mu_{e}$. As we can see in Fig.~\ref{fig:Yfrac_DD-ME2} and Fig.~\ref{fig:chem_PKDD}, the total hyperon fraction $X_{Y}$ is higher for the $npe\mu Y$ matter and the fraction of the neutron is suppressed so that the Fermi momentum of the neutron and $\mu_{n}+\mu_{e}$~rise slowly. For the results calculated with PKDD, $\mu_{\Sigma^{-}}$ and $\mu_{n}+\mu_{e}$ are approximately parallel after $\Xi^{-}$ is populated. They will not intersect as density increases and do not satisfy the threshold equation. However, $X_{Y}$ is suppressed when~$\phi$-meson is considered. The Fermi momentum of the neutron is higher and $\mu_{n}+\mu_{e}$ rises faster than the case without~$\phi$-meson. The result is that $\mu_{n}+\mu_{e}$ is easier to intersect with $\mu_{\Sigma^{-}}$, causing $\Sigma^{-}$ to appear in the $npe\mu Y\phi$ matter. The changed hyperon species inside neutron stars may affect the cooling properties induced by the hyperons. Similar discussions hold also in the calculation results with DDVT and DDVTD.
\begin{figure}[ht!]
\centering
\includegraphics[scale=0.48]{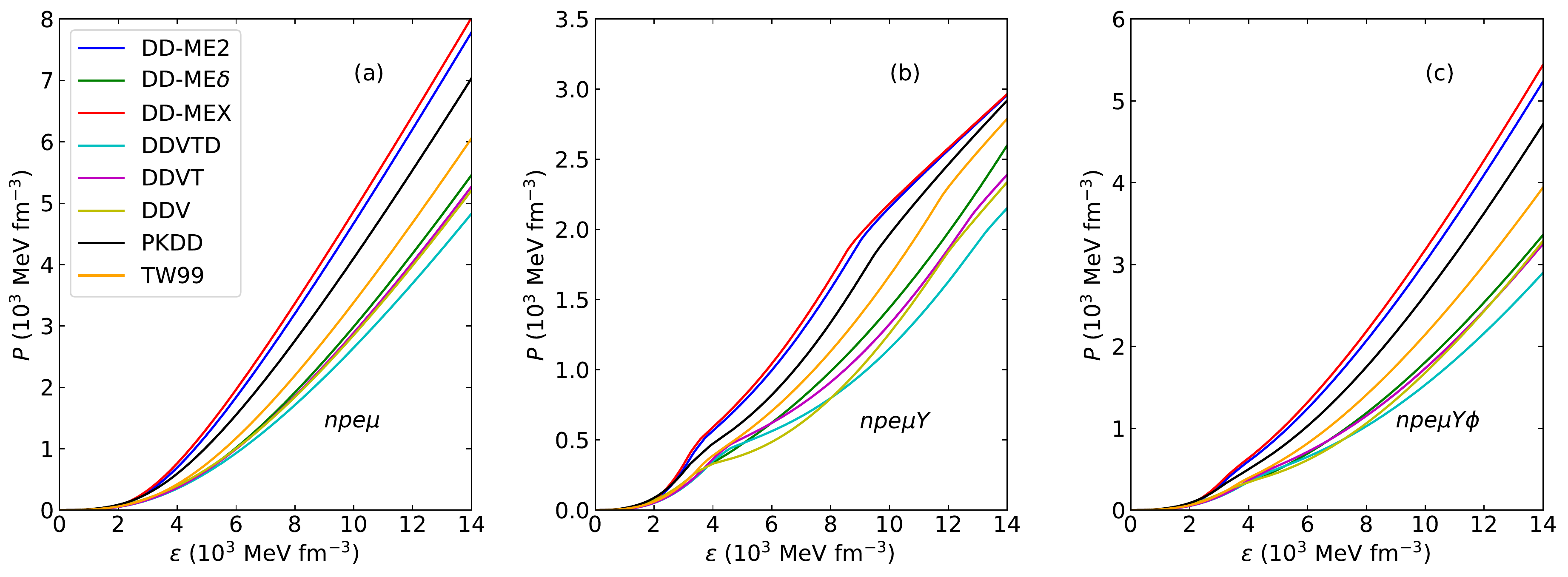}
\caption{EoSs of $npe\mu$, $npe\mu Y$, and $npe\mu Y\phi$ matters with various effective interactions. \label{fig:EoS}}
\end{figure}
\subsection{Effects of $\phi$-meson on the EoS and sound velocity} \label{subsec:phi_on_EoS_vs}
The $\phi$-meson significantly changes the neutron star interior composition and affects the EoS and the sound velocity~$v_{s}$. In Fig.~\ref{fig:EoS}, the EoSs of~$npe\mu$, $npe\mu Y$, and~$npe\mu Y\phi$~matters calculated with various effective interactions are displayed. For the $npe\mu$ matter, the two stiffest EoSs are given by DD-ME2 and DD-MEX and the softest EoS is given by DDVTD. Although the saturation properties of nuclear matter of DDV and DDVT are very different, the EoSs given by them are very close to each other even at high energy density. Comparing the EoSs generated with DDVT and DDVTD, the~$\delta$-meson softens the EoS of asymmetric nuclear matter in the DDRMF model, which is consistent  with previous studies~\citep{Liu2007_PRC75-048801,Liu2008_CTP49-199,Wang2014_PRC90-055801}.

The appearance of hyperons in the $npe\mu Y$ matter leads to a strong softening of the EoSs for all effective interactions. The DD-ME2 and DD-MEX still give the two stiffest EoSs. The pressure calculated with DD-MEX is obviously larger than that calculated with DD-ME2 at low energy density but there is little difference at high energy density. The softest EoS is given by DDV at low energy density but DDVTD generates the softest EoS at high energy density. The EoS calculated with DDVT is significantly stiffer than that calculated with DDV at low energy density compared to the $npe\mu$ matter. For all effective interactions, the EoSs of the $npe\mu Y\phi$ matter are stiffer than those of the $npe\mu Y$ matter, but they are still softer than those of the $npe\mu$ matter. The $\phi$-meson shifts the hyperon thresholds to higher densities and suppresses the hyperon fractions, causing a weaker softening of the total EoS.
\begin{figure}[ht!]
\centering
\includegraphics[scale=0.47]{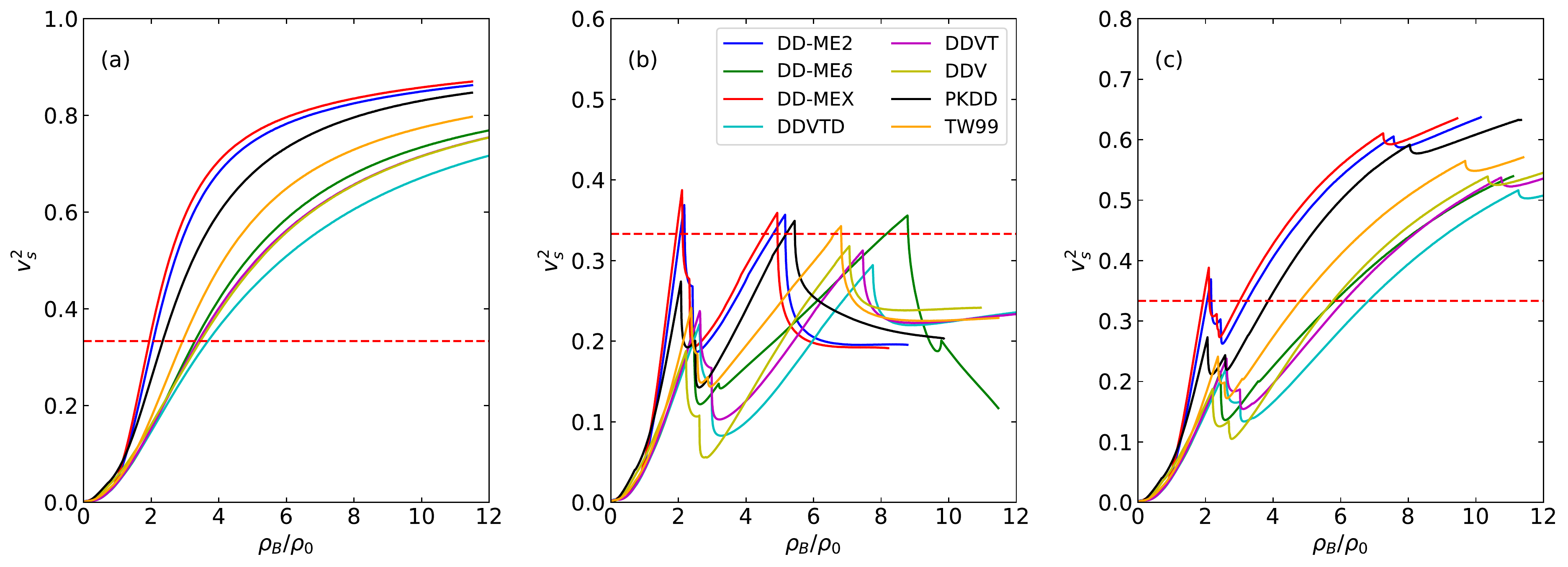}
\caption{Squared sound velocity as a function of $\rho_{B}/\rho_{0}$ for $npe\mu$, $npe\mu Y,$ and $npe\mu Y\phi$ matters with various effective interactions. The dashed line represents the conformal limit~$v_{s}^{2}=1/3$.  \label{fig:SoundVelocity}}
\end{figure}

The squared sound velocity $v_{s}^{2}$ can be easily obtained from the EoS by using Eq.~(\ref{equ:DDRMF_soundvelocity}). The causal limit~$v_{s}^{2}<1$~\citep{Dutra2016_PRC93-025806}~and the conformal limit~$v_{s}^{2}<1/3$~\citep{Bedaque2015_PRL114-031103}~are used to constrain sound velocity. For the $npe\mu$ matter, the sound velocities calculated with all effective interactions satisfy the causal limit, but they exceed the conformal limit. A stiffer EoS leads to a higher sound velocity. For the $npe\mu Y$ matter, the sound velocity is reduced because hyperons soften the corresponding EoS. The causal limit is also fulfilled for all effective interactions, but the squared sound velocities exceed the conformal limit around some hyperon thresholds and are less than one-third at high density. We also notice that the sound velocity shows a peak at the threshold density of every individual hyperon because the onset of hyperons suddenly softens the EoS~\citep{Lopes2014_PRC89-025805}. Unlike the $npe\mu$ matter, it seems impossible to draw the conclusion that the stiffer EoS leads to a larger sound velocity, e.g., the sound velocity calculated with PKDD is smaller than that of DDVTD, but the EoS calculated with PKDD is significantly stiffer than that calculated with DDVTD. For the $npe\mu Y\phi$ matter, the sound velocities calculated with all effective interactions satisfy the causal limit, but the conformal limit is exceeded at high density. The $\phi$-meson suppresses the hyperon fractions of the $npe\mu Y\phi$ matter and weakens the rapid reduction of sound velocity caused by the softening of related EoS. Meanwhile, the increased hyperon thresholds make the sound velocity increase again in a larger density range, which leads to the conformal limit being exceeded.
\begin{figure}[ht!]
\centering
\includegraphics[scale=0.5]{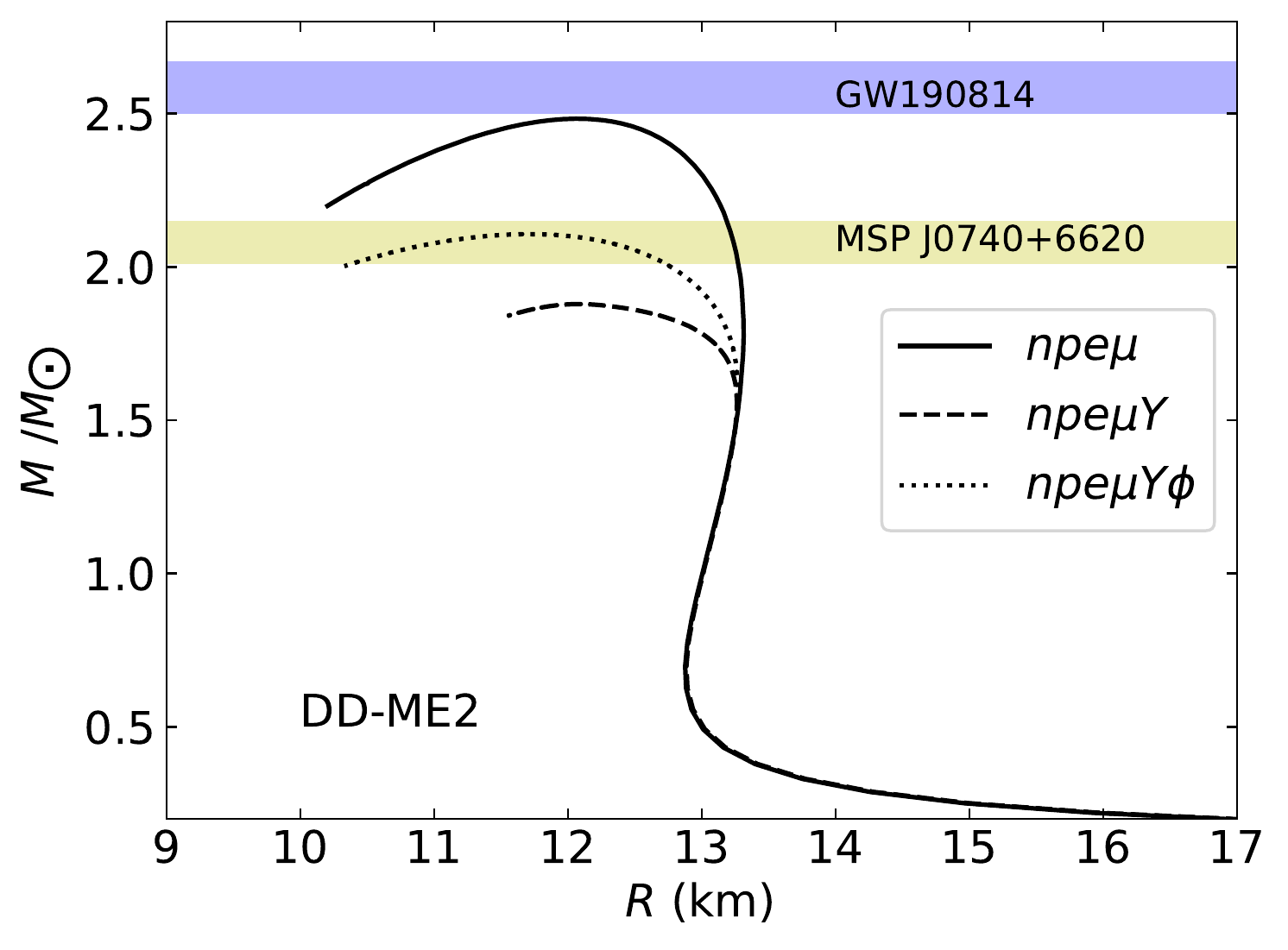}
\caption{Mass--radius relations of~$npe\mu$, $npe\mu Y$, and~$npe\mu Y\phi$~neutron stars calculated with DD-ME2. \label{fig:MR_DD-ME2}}
\end{figure}
\subsection{Effects of $\phi$-meson on the mass--radius relation} \label{subsec:phi_on_MR}
The $\phi$-meson has an impact on the macroscopic properties of neutron stars which are closely related to the EoS. The mass--radius relation of a static neutron star is obtained by solving the TOV equation (\ref{equ:TOV})~with a given EoS as input. The BPS~\citep{Baym1971_ApJ170-299}~and~BBP~\citep{Baym1971_NPA175-225}~EoSs are chosen as the EoSs of outer and inner crust of neutron stars, respectively. Taking DD-ME2 as an example, the mass--radius relations of neutron stars are shown in Fig.~\ref{fig:MR_DD-ME2}. The largest maximum mass of neutron stars is obtained by using the EoS without hyperons. The smallest maximum mass of neutron stars is obtained by using the EoS with hyperons but without $\phi$-meson. The maximum mass of $npe\mu Y\phi$ neutron stars is larger than that of $npe\mu Y$ neutron stars but smaller than that of $npe\mu$ neutron stars because the stiffness of the EoS with hyperons and $\phi$-meson is between that of the EoS with hyperons but without $\phi$-meson and that of the EoS without hyperons. Results calculated with other effective interactions show similar characteristics to that calculated with DD-ME2.
\begin{figure}[ht!]
\centering
\includegraphics[scale=0.45]{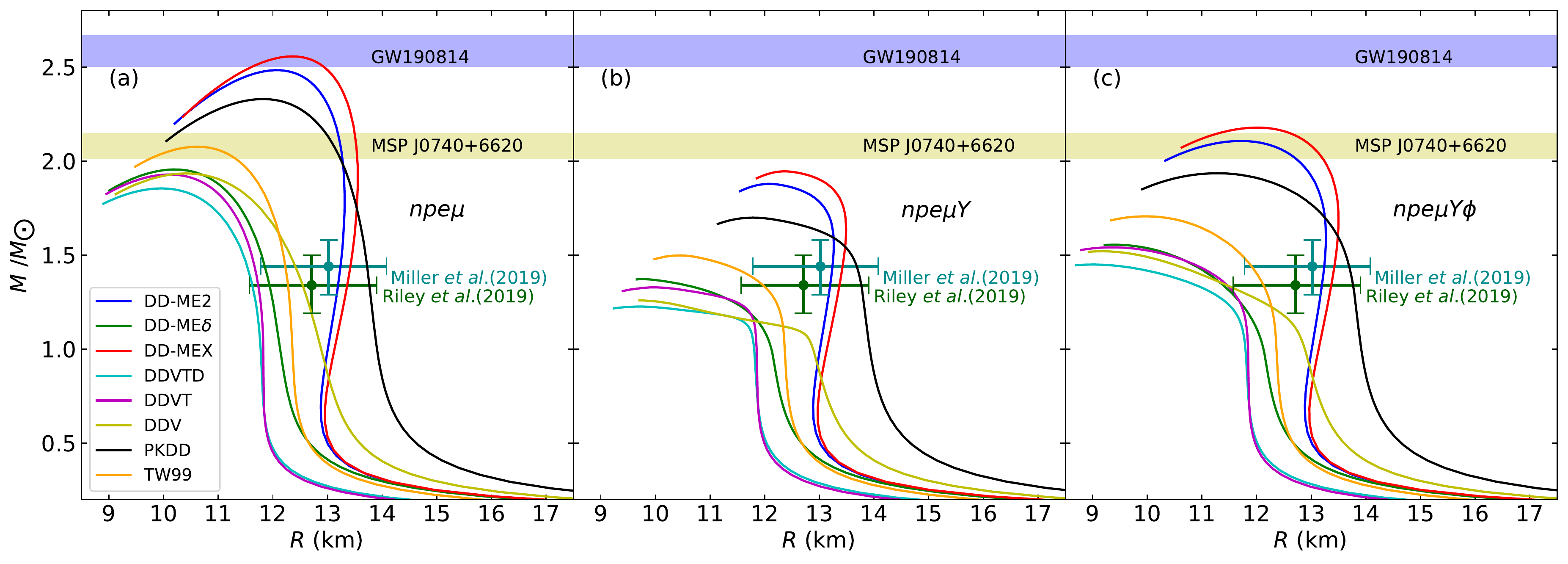}
\caption{Mass--radius relations of~$npe\mu$, $npe\mu Y$, and~$npe\mu Y\phi$~neutron stars calculated with various effective interactions. The constraints from astrophysical observables of MSP J0740+6620~(yellow area), the secondary object of GW190814~(blue area), and the mass and radius of PSR J0030+0451 from NICER in~\citet{Riley2019_ApJL887-L21}~(dark green error bar) and~\citet{Miller2019_ApJL887-L24}~(dark cyan error bar) are shown. \label{fig:MR}}
\end{figure}
\begin{deluxetable*}{lrrrrrrrrrrrr}
\tablenum{4}
\tablecaption{Properties of $npe\mu$, $npe\mu Y$, and $npe\mu Y\phi$ neutron stars calculated with various effective interactions.\label{tab:MR}}
\tablewidth{2pt}
\tablehead{
\colhead{Effective Interaction} & \colhead{$M_{\mathrm{max}}$} & \colhead{$R$} & \colhead{$\rho_{c}$} & \colhead{$R_{1.4M_{\odot}}$}
                    & \colhead{$M_{\mathrm{max}}$} & \colhead{$R$} & \colhead{$\rho_{c}$} & \colhead{$R_{1.4M_{\odot}}$}
                    & \colhead{$M_{\mathrm{max}}$} & \colhead{$R$} & \colhead{$\rho_{c}$} & \colhead{$R_{1.4M_{\odot}}$} \\
                    & \colhead{($M_{\odot}$)} & \colhead{(km)} & \colhead{(fm$^{-3}$)} & \colhead{(km)}
                    & \colhead{($M_{\odot}$)} & \colhead{(km)} & \colhead{(fm$^{-3}$)} & \colhead{(km)}
                    & \colhead{($M_{\odot}$)} & \colhead{(km)} & \colhead{(fm$^{-3}$)} & \colhead{(km)} \\
                    & \multicolumn{4}{c}{$npe\mu$} & \multicolumn{4}{c}{$npe\mu Y$} & \multicolumn{4}{c}{$npe\mu Y\phi$} \\
                    \cmidrule(r){2-5} \cmidrule(r){6-9} \cmidrule(r){10-13}
}
\decimalcolnumbers
\startdata
DD-ME2        & 2.483 & 12.060 & 0.82 & 13.237 & 1.879 & 12.081 & 0.90 & 13.240 & 2.107 & 11.719 & 0.93 & 13.239 \\
DD-ME$\delta$ & 1.955 & 10.207 & 1.21 & 11.852 & 1.373 & 9.769  & 1.55 & ---    & 1.555 & 9.386  & 1.59 & 10.896 \\
PKDD          & 2.330 & 11.808 & 0.89 & 13.725 & 1.699 & 11.781 & 1.02 & 13.684 & 1.935 & 11.297 & 1.06 & 13.695 \\
TW99          & 2.076 & 10.615 & 1.10 & 12.228 & 1.498 & 10.447 & 1.29 & 11.369 & 1.706 & 9.980  & 1.36 & 11.911 \\
DD-MEX        & 2.556 & 12.374 & 0.77 & 13.419 & 1.945 & 12.364 & 0.85 & 13.424 & 2.178 & 12.010 & 0.88 & 13.424 \\
DDV           & 1.934 & 10.405 & 1.20 & 12.409 & $>$1.259 & $<$9.706 & $>$1.65 & ---  & 1.521 & 9.177  & 1.71 & 10.588 \\
DDVT          & 1.929 & 10.111 & 1.23 & 11.664 & 1.329 & 9.975  & 1.44 & ---    & 1.541 & 9.361  & 1.59 & 10.944 \\
DDVTD         & 1.855 & 9.953  & 1.29 & 11.529 & 1.227 & 9.714  & 1.56 & ---    & 1.450 & 8.992  & 1.77 & 9.537  \\
\enddata
\tablecomments{$M_{\mathrm{max}}$, $R$, $\rho_{c}$, and $R_{1.4M_{\odot}}$ denote maximum mass, radius corresponding to maximum mass, central density, and radius of neutron star at 1.4$M_{\odot}$, respectively.}
\end{deluxetable*}

In Fig.~\ref{fig:MR}, the mass--radius relations of neutron stars calculated with various effective interactions for $npe\mu$, $npe\mu Y$, and $npe\mu Y\phi$ neutron stars are shown. The corresponding neutron star properties are listed in Table~\ref{tab:MR}. Based on chiral effective field theory interactions, a radius constraint on the canonical 1.4~$M_{\odot}$~neutron star with~9.7~km $ \leqslant R_{1.4M_{\odot}} \leqslant13.9$ km was given in~\citet{Hebeler2010_PRL105-161102}. From Table~\ref{tab:MR}, most of our results are compatible with this prediction. The mass of PSR J0030+0451 from NICER is close to the canonical neutron star mass, as shown in the dark green and dark cyan error bars in Fig.~\ref{fig:MR}. It can be found that the mass--radius relations of~$npe\mu$ neutron stars calculated with all effective interactions agree well with the observations from NICER. But DD-ME$\delta$, DDV, DDVT, and DDVTD are excluded by the constraints from NICER, both for $npe\mu Y$ neutron stars and for $npe\mu Y \phi$ neutron stars. The $\phi$-meson does not seem to significantly affect the radius constraint.

There exists a strong tension between the existence of neutron stars with mass around 2 $M_{\odot}$ and the conformal limit~\citep{Bedaque2015_PRL114-031103}. The large sound velocity that violates the conformal limit may appear inside neutron star cores with the constraints of astrophysical observations~\citep{Tews2018_ApJ860-149,Reed2020_PRC101-045803}. Alsing~\emph{et al}. \citep{Alsing2018_MNRAS478-1377}~found the lower bound on the maximum sound velocity of~$v_{s}^{\mathrm{max}}>0.63$~inside neutron stars so that the conformal limit is broken significantly. However, Ma and Rho~\citep{Ma2019_PRD100-114003}~developed the pseudo-conformal model and found that the maximum mass constraint of~2.3 $M_{\odot}$~is accommodated by setting in the conformal limit at~$\rho_{B}\gtrsim 2\rho_{0}$. In the present work, although $\phi$-meson decreases the sound velocity inside neutron stars, the tension mentioned above has not been resolved in the DDRMF model. In Table~\ref{tab:MR}~and Fig.~\ref{fig:SoundVelocity}, we can see that the conformal limit is broken in neutron stars whose maximum masses reach~2 $M_{\odot}$.

The MSP J0740+6620 with $2.08_{-0.07}^{+0.07} M_{\odot}$~(68.3\% credibility interval)~\citep{Fonseca2021_arXiv2104.00880} is used to constrain the maximum mass of neutron stars in the present work. From Fig.~\ref{fig:MR}~and Table~\ref{tab:MR}, MSP J0740+6620~can rule out DD-ME$\delta$, DDV, DDVT, and DDVTD for~$npe\mu$~neutron stars; for~$npe\mu Y$~neutron stars, all effective interactions are ruled out because none of them gives a maximum mass larger than the lower bound of mass of MSP J0740+6620; for~$npe\mu Y\phi$~neutron stars, the maximum masses calculated with DD-ME2 and DD-MEX are compatible with the observed mass range of MSP J0740+6620, while other effective interactions are excluded. Although~$\phi$-meson increases the maximum mass of neutron stars, only a few effective interactions survive under the constraints of astrophysical observations.

The GW190814 event, which is a compact binary merger involving a 22.2--24.3 $M_{\odot}$ black hole and a compact object of 2.50--2.67 $M_{\odot}$, was detected by the LIGO/Virgo in August of~2019~\citep{Abbott2020_ApJL896-L44}. Whether the secondary object of GW190814~is a massive neutron star or a low mass black hole (BH) remains controversial. Huang \emph{et al}. \citep{Huang2020_ApJ904-39} suggested that the possibility of the secondary object of GW190814~as a neutron star consisting of hadron matter is not excluded in the DDRMF model. Zhang and Li \citep{Zhang2020_ApJ902-38} found that the secondary object of GW190814~could be a massive pulsar with the highest rotational frequency ever observed, and several following studies indicated that this pulsar may have a quark or hyperonic core \citep{Dexheimer2021_PRC103-025808,Rather2021_PRC103-055814}. The $R$-mode stability of this superfast pulsar is supported by \citet{Zhou2021_ApJ910-62}. The possibility of the secondary object of GW190814~as a hyperon star or a low mass black hole was studied in~\citet{Sedrakian2020_PRD102-041301R}~and~\citet{Li2020_PLB810-135812} and it was implied that the GW190814 event was likely to be binary BH merger rather than NS--BH merger. Recently, the possibilities of strange quark star \citep{Bombaci2021_PRL126-162702}, up-down quark star \citep{Cao2020_arXiv2009.00942}, and dark matter admixed neutron star \citep{Das2021_PRD104-063028} were also proposed. In the present work, the DD-ME2 and DD-MEX support that the secondary object of GW190814~is a neutron star without hyperons, similar to the conclusion drawn in~\citet{Huang2020_ApJ904-39}. However, there are no effective interactions supporting it as a hyperon star whether or not $\phi$-meson is considered.

\begin{figure}[ht!]
\centering
\includegraphics[scale=0.45]{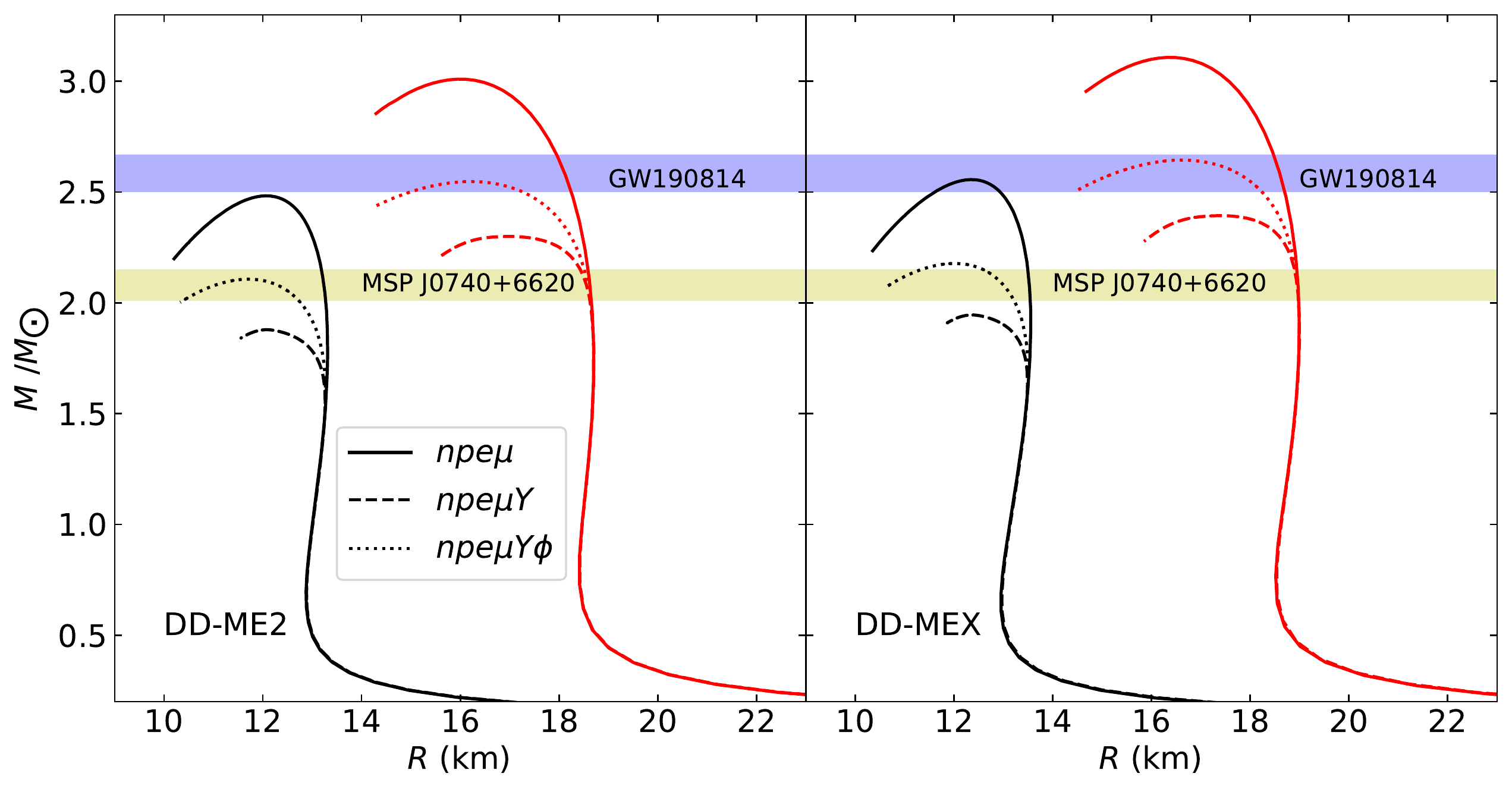}
\caption{Mass--radius relations of non-rotating~(black lines)~and Keplerian rotating~(red lines)~neutron stars calculated with DD-ME2 and DD-MEX for $npe\mu$, $npe\mu Y$, and $npe\mu Y\phi$ neutron stars. \label{fig:SRNSMR}}
\end{figure}

As is well known, the rotation of neutron stars can increases the maximum mass by about 20\% \citep{Weber1992_ApJ390-541,Cook1994_ApJ424-823,Paschalidis2017_LRR20-7}. For the static counterpart of the secondary object in GW190814, the lower limit on its maximum mass is around 2.08 $M_{\odot}$~\citep{Most2020_MNRASL499-L82}. From Table~\ref{tab:MR}, we note that the maximum masses of~$npe\mu Y\phi$~neutron stars calculated with DD-ME2 and DD-MEX exceed 2.1 $M_{\odot}$ and hence the two effective interactions support the possibility of the secondary object of GW190814 as a rotating~$npe\mu Y\phi$~neutron star. In order to verify this possibility, using the previous EoSs calculated with DD-ME2 and DD-MEX, it is necessary to calculate the global properties of neutron stars rotating with the Kepler frequency. For a uniformly rotating neutron star with axisymmetric configuration, the metric tensor, which describes its geometry, is given by
\begin{equation}\label{equ:RNS_metric}
\begin{aligned}
  \mathrm{d}s^{2}=-e^{2\nu(r,\theta)}\mathrm{d}t^{2}
                  +e^{2\psi(r,\theta)}[ \mathrm{d}\phi-\omega(r)\mathrm{d}t ]
                  +e^{2\mu(r,\theta)}\mathrm{d}\theta^{2}
                  +e^{2\lambda(r,\theta)}\mathrm{d}r^{2},
\end{aligned}
\end{equation}
where the gravitational potentials $\nu$, $\psi$, $\mu$, and $\lambda$ depend on the radial variable $r$ and the azimuthal angle $\theta$ but independent of time $t$ and the polar angle $\phi$ \citep{Butterworth1976_ApJ204-200,Friedman1986_ApJ304-115}. The RNS code \citep{Cook1994_ApJ424-823, Stergioulas1995_ApJ444-306}, which is based on the Komatsu-Eriguchi-Hachisu (KEH) method \citep{Komatsu1989_MNRAS237-355}, iteratively solves the Einstein's field equations and the hydrostatic equilibrium equation and give the numerical solution of the equilibrium structure of a rotating neutron star. The rotation is taken into account by deforming a neutron star from spherical configuration to axisymmetric configuration. The Kepler frequency of a stable neutron star is the maximum spin frequency above which the mass escapes from its surface \citep{Rather2021_PRC103-055814}. The surface gravitational redshift, which is related to the compactness $M/R$, can be used to constrain the EoSs of neutron stars. The surface gravitational redshift $z_{\mathrm{surf}}$ of the canonical static neutron star is calculated as $z_{\mathrm{surf}}=(1-2M/R)^{-1/2}-1$ \citep{Tolos2017_PASA34-e065}. For a rapidly rotating neutron star, its polar redshift $z_{p}$, equatorial redshift in the backward direction $z_{b}$, and quatorial redshift in the forward direction $z_{f}$ are obtained by using the equations taken from APPENDIX B in \citet{Cook1994_ApJ424-823}.

In Fig.~\ref{fig:SRNSMR}, we display the mass--radius relations of static and Keplerian rotating neutron stars with DD-ME2 and DD-MEX. The maximum masses of neutron stars at Kepler frequency are larger than their static counterparts. A rotating neutron star has a larger equatorial radius than its static counterpart for both the canonical stars and neutron stars with maximum mass. The properties of static and Keplerian rotating neutron stars calculated with DD-ME2 and DD-MEX are given for comparison in Table \ref{tab:RNSMR}. The rotations of neutron stars increase their maximum masses by 20.9\%--23.1\%. The gravitational redshifts of $npe\mu$, $npe\mu Y$, and $npe\mu Y\phi$ neutron stars are almost the same for DD-ME2 and DD-MEX because the effects of hyperons and $\phi$-meson are weak inside the canonical neutron star cores. Hebeler \emph{et al}. \citep{Hebeler2010_PRL105-161102} gave a gravitational redshift range of $z=0.193$--$0.320$ and an observational limit of $z=0.12$--$0.23$ from 1E 1207.4-5209 was presented in \citet{Sanwal2002_ApJ574-L61}. From our results, both $z_{\mathrm{surf}}$ of static neutron stars and the polar redshift $z_{p}$ of Keplerian rotating neutron stars match the values from space telescopes \citep{Douchin2001_AA380-151,Sanwal2002_ApJ574-L61}. From the maximum masses listed in Table \ref{tab:RNSMR}, DD-ME2 and DD-MEX support that GW190814's secondary object is a hyperon star spinning faster than 1264 Hz and 1170 Hz, respectively, but an important prerequisite is that the $\phi$-meson should be included.

\begin{deluxetable*}{lrrrrrrrrrrrr}

\tablenum{5}
\tablecaption{Properties of static and Keplerian rotating $npe\mu$, $npe\mu Y$, and $npe\mu Y\phi$ neutron stars calculated with DD-ME2 and DD-MEX.\label{tab:RNSMR}}
\tablewidth{2pt}
\tablehead{
\colhead{      } & \multicolumn{6}{c}{Static NS} & \multicolumn{6}{c}{Keplerian rotating NS} \\
\cmidrule(r){2-7} \cmidrule(r){8-13}
\colhead{      } & \multicolumn{3}{c}{DD-ME2}    & \multicolumn{3}{c}{DD-MEX} & \multicolumn{3}{c}{DD-ME2}    & \multicolumn{3}{c}{DD-MEX}\\
\cmidrule(r){2-4} \cmidrule(r){5-7} \cmidrule(r){8-10} \cmidrule(r){11-13}
\colhead{      } & \colhead{$npe\mu$} & \colhead{$npe\mu Y$} & \colhead{$npe\mu Y\phi$} & \colhead{$npe\mu$} & \colhead{$npe\mu Y$} & \colhead{$npe\mu Y\phi$} & \colhead{$npe\mu$} & \colhead{$npe\mu Y$} & \colhead{$npe\mu Y\phi$} & \colhead{$npe\mu$} & \colhead{$npe\mu Y$} & \colhead{$npe\mu Y\phi$}
}
\decimalcolnumbers
\startdata
$M_{\mathrm{max}}$ ($M_{\odot}$)      & 2.483 & 1.879 & 2.107 & 2.556 & 1.945 & 2.178 & 3.010 & 2.300 & 2.548 & 3.108 & 2.394 & 2.645 \\
$R$ (km)               & 12.060& 12.081& 11.719& 12.374& 12.364& 12.010& 16.005& 16.989& 16.215& 16.377& 17.402& 16.618\\
$\rho_{c}$ (fm$^{-3}$) & 0.82  & 0.90  & 0.93  & 0.77  & 0.85  & 0.88  & 0.73  & 0.73  & 0.79  & 0.69  & 0.69  & 0.74  \\
$R_{1.4M_{\odot}}$ (km)& 13.237& 13.240& 13.239& 13.419& 13.424& 13.424& 18.640& 18.640&18.640 & 18.859& 18.862& 18.861\\
$z_{\mathrm{surf}}$    & 0.206 & 0.206 & 0.206 & 0.202 & 0.202 & 0.202 & ---   & ---   & ---   & ---   & ---   & ---   \\
$z_{p}$                & ---   & ---   & ---   & ---   & ---   & ---   & 0.221 & 0.221 & 0.221 & 0.218 & 0.217 & 0.217 \\
$z_{f}$                & ---   & ---   & ---   & ---   & ---   & ---   & $-$0.238& $-$0.238& $-$0.238& $-$0.237& $-$0.237& $-$0.237\\
$z_{b}$                & ---   & ---   & ---   & ---   & ---   & ---   & 0.700 & 0.700 & 0.700 & 0.692 & 0.691 & 0.691 \\
\enddata
\tablecomments{The definitions of $M_{\mathrm{max}}$ and $\rho_{c}$ are the same as those in Table \ref{tab:MR}. $R$ and $R_{1.4 M_{\odot}}$ are the equatorial radii of the neutron star with maximum mass and canonical neutron star, respectively. $z_{\mathrm{aurf}}$ is the surface gravitational redshift of static neutron star. $z_{p}$, $z_{f}$, and $z_{b}$ are the polar redshift, the equatorial redshift in the forward direction, and the equatorial redshift in the backward direction of rapidly rotating neutron star, respectively.}
\end{deluxetable*}

\section{Summary} \label{sec:summar}
The effects of~$\phi$-meson on the properties of hyperon stars have been studied systematically in the DDRMF model. The widely used (DD-ME2, DD-ME$\delta$, PKDD, and TW99) and the latest proposed (DD-MEX, DDV, DDVT, and DDVTD) effective interactions were applied to calculate the interior compositions, equations of state, sound velocities, and mass--radius relations of neutron stars.

Similar to the pervious works \citep{Banik2014_ApJSupp214-22,Biswal2019_ApJ885-25,Lopes2020_EPJA56-122}, since the $\phi$-meson mediates the repulsive interaction between hyperons, the hyperon thresholds are shifted to higher densities and the total hyperon fraction decreases when $\phi$-meson is included. The $\phi$-meson has a greater impact on~$\Xi$-hyperon because its strangeness number is $-2$ and the coupling strength between $\Xi$ and $\phi$ is twice of those for other hyperons. The reversal of the order of hyperon thresholds and the emergence of $\Sigma^{-}$ in the $npe\mu Y\phi$ matter are explained by the effects of $\phi$-meson on hypernuclear matter.

Because the hyperon composition is suppressed in the $npe\mu Y\phi$ matter, the $\phi$-meson significantly stiffens the EoSs and causes the sound velocity to exceed the conformal limit. The tension between the conformal limit and the existence of neutron stars with 2~$M_{\odot}$~ still exists in this work, because the conformal limit is broken in neutron stars whose maximum masses reach~2~$M_{\odot}$.

The~$\phi$-meson increases the maximum mass of neutron stars due to the stiffening of corresponding EoSs in the DDRMF model, which is consistent with earlier conclusions \citep{Weissenborn2012_NPA881-62,Banik2014_ApJSupp214-22,Biswal2019_ApJ885-25,Lopes2020_EPJA56-122}. For neutron stars with hyperon but without~$\phi$-meson, the mass of PSR J0740+6620~rules out all effective interactions used in the present work. However, the DD-ME2 and DD-MEX survive under the mass constraint from PSR J0740+6620 when~$\phi$-meson is included. For the secondary object of GW190814, whether or not to consider $\phi$-meson, our results do not support it as a hyperon star, but two effective interactions, i.e., DD-ME2 and DD-MEX, support that it is an $npe\mu Y\phi$ neutron star rotating with Kepler frequency.

\begin{acknowledgments}
Helpful discussions with Johann Haidenbauer, Hoai Le and
Andreas Nogga are gratefully acknowledged. We thank Xiang-Xiang Sun and Yu-Ting Rong for reading the manuscript and valueable suggestions.
This work has been support by the National Key R\&D Program of China (Grant No. 2018YFA0404402),
the National Natural Science Foundation of China (Grants
No. 11525524, No. 12070131001, No. 12047503, and No. 11961141004),
the Key Research Program of Frontier Sciences of Chinese Academy of Sciences (Grant No. QYZDB-SSWSYS013), the Strategic Priority Research Program of Chinese Academy of Sciences (Grants No. XDB34010000 and No. XDPB15), and the IAEA Coordinated Research Project (Grant No. F41033). The results described in this paper are obtained on the High-performance Computing Cluster of ITP-CAS and the ScGrid of the Supercomputing Center, Computer Network Information Center of Chinese Academy of Sciences.
\end{acknowledgments}

\bibliography{zhtu}{}
\bibliographystyle{aasjournal}



\end{document}